\documentclass[published]{JHEP3}

\JHEP{ }

\usepackage{epsf}
\usepackage{latexsym}
\usepackage{epsfig,multicol}

\def\yzero{\smash{\hbox{$y\kern-4pt\raise1pt\hbox{${}^\circ$}$}}}
\def\p{\partial}
\def\a{\alpha}

\def\d{\delta}
\def\beq{\begin{equation}}
\def\eeq{\end{equation}}
\def\beqa{\begin{eqnarray}}
\def\eeqa{\end{eqnarray}}
\def\Om{\Omega}

\def\vt{\vartheta}

\def\-{\hphantom{-}}
\def\ov{\overline}
\def\s2{\frac{1}{\sqrt2}}

\def\oh{\frac{1}{2}}
\def\beq{\begin{equation}}
\def\eeq{\end{equation}}
\def\beqa{\begin{eqnarray}}
\def\eeqa{\end{eqnarray}}

\def\IF{\relax{\rm I\kern-.18em F}}
\def\II{\relax{\rm I\kern-.18em I}}
\def\IP{\relax{\rm I\kern-.18em P}}
\def\IC{\relax\hbox{\kern.25em$\inbar\kern-.3em{\rm C}$}}
\def\IR{\relax{\rm I\kern-.18em R}}

\def\ck{{\cal K}}

\def\cn{{\cal N}}
\def\cam{{\cal M}}

\def\cf{{\cal F}}

\def\Dsl{\,\raise.15ex\hbox{/}\mkern-13.5mu D} 
\def\IZ{Z\kern-.4em  Z}

\def\t{\times}
\def\eps{\epsilon}
\def\A{\Arrowvert}
\def\car{{\cal R}}
\def\OR{\Omega {\cal R}}
\def\ent{{\bf Z}}

\newbox\pippobox

\title{SUSY Quivers, Intersecting Branes and the Modest Hierarchy Problem}

\author{D.~Cremades, L.~E.~Ib\'a\~nez and  F.~Marchesano \\
 	Departamento de F\'{\i}sica Te\'orica C-XI
	and Instituto de F\'{\i}sica Te\'orica  C-XVI,\\
	Universidad Aut\'onoma de Madrid,
	Cantoblanco, 28049 Madrid, Spain.
}
\received{\today}               
\revised{\today}
\accepted{}               

\preprint{\hepth{0201205}}
\preprint{FTUAM-02/3 IFT-UAM/CSIC-02-1}

\abstract{We present a class of chiral non-supersymmetric $D=4$ field
theories in which quadratic divergences appear only at two loops.
They may be depicted  as ``SUSY quivers''  in which the nodes
represent a gauge group with extended e.g., $N=4$  SUSY
whereas links represent bifundamental matter fields
which transform as chiral multiplets with respect to
{\it different}  $N=1$ subgroups.
One can obtain this type of field theories from simple D6-brane
configurations on Type IIA string theory compactified on a six-torus.
We discuss the conditions under which this kind of structure
is obtained from D6-brane intersections.
 We also discuss some
aspects of the effective low-energy field theory.
In particular we  compute gauge couplings and Fayet-Iliopoulos
terms from the Born-Infeld  action and show
how they match  the  field theory results.
This class of theories may be  of
phenomenological interest in order to understand the
{\it modest hierarchy problem} i.e., the stability of the
hierarchy between the weak scale and a fundamental scale of order
10-100 TeV which appears e.g. in low string scale models.
Specific D-brane models with the spectrum of the SUSY Standard
Model and three generations are presented.}


\begin{document}

\section{Introduction}

One of the most relevant properties of Dp-branes 
\cite{poldb} in string theory is
the fact that they localize gauge (and matter) interactions on a
(p+1)-dimensional submanifold of the full higher dimensional  
space-time. This property opens the way to 
the generation of ``mixed symmetry field theories'' with different
subsectors (coming from different localized sectors) with
different symmetries and supersymmetries. Thus, depending
from which sector of a given brane configuration
a field theory is originated, one gets different
symmetries and properties.
One can envisage, for example,  a field theory in which the pure gauge
sector has $\cn=4$ supersymmetry and some other massless fields
respect only a subgroup or none of those supersymmetries.
Whereas writing such type of Lagrangians from scratch
would look sort of contrived, from the string-theory and
D-brane point of view they  turn out to appear naturally.

Supersymmetry ($\cn=1$) has been applied to phenomenology in the last
two decades due to the fact that quadratic divergences
which may destabilize the hierarchy of scales are
canceled. It is of obvious interest the
search for $\cn=0$ theories in which there is some level
of suppression of quadratic divergences.
In particular, it has been recently realized
that the scale of fundamental physics could be anywhere between, say
1-TeV and the Planck scale $M_p$
\cite{uniwitten,lykken,untev,anton} 
. The largeness of the Planck
scale would then be an artifact of the presence of large extra dimensions
or warp factors of the metric \cite{rs}.
 Thus it has been put forward the idea that the
weak scale could be associated to the string scale
\cite{lykken,untev,anton}
. On the other hand
the SM works so well that it is difficult to make the string scale
lighter than a few TeV without entering into conflict with experimental
data (see e.g. ref.\cite{barbieri} for a nice physical view
of the problem).  Furthermore, the
absence of any exotic source of flavour
changing neutral currents (FCNC) would be most easily guaranteed if
the string scale was postponed to scales of order 10-100 TeV.
But if one has $M_s\sim 10-100$ TeV, we will have to explain the
relative smallness of the weak scale $M_Z=90$ GeV compared to the
string scale. This we call the ``modest hierarchy problem''.

An obvious cure to this problem is the full suppression of quadratic
divergences offered by standard $\cn  = 1$ SUSY theories. On the other hand
that is much more than what we actually need. To make a mass hierarchy of
three orders of magnitude sufficiently stable it is enough to have
absence of quadratic divergences up to  one-loop, not at all loops.
In the present paper, we present a class of
chiral non-supersymmetric D=4 theories in which quadratic divergences only
appear at two loops. We will
call these models "quasi-supersymmetric"
(Q-SUSY)
because different subsectors of the theory respect different $\cn=1$  
supersymmetries, but the complete Lagrangian has $\cn=0$
\footnote{It would be appropriate (particularly from
the brane point of view)  to call these theories
{\it locally supersymmetric} but this may lead to confusion
with gauged supersymmetry, i.e., supergravity.}.
We think that this class of theories are interesting in their own
right. Furthermore they offer a solution to the
``modest hierarchy problem'' described above. Absence of quadratic divergences
up to one-loop would be enough to maintain a hierarchy between  a string
scale of order 10-100 TeV and the weak scale.
This class of Q-SUSY gauge theories may be naturally obtained
in a string context.
One can show explicit  string constructions
with intersecting \cite{bdl} 
D-branes wrapping a 6-torus  \cite{bgkl,afiru,bkl,imr}
giving    rise to such type
field theories as their low-energy limit.   
We will be in fact able to construct concrete D6-brane
models with a massless content quite analogous to that of the
Minimal Supersymmetric Standard Model.

The class of models that we discuss in detail have a pure
gauge sector respecting $\cn=4$ supersymmetry with a gauge
group $\prod _{i} U(M_i)$. In addition there are
scalars and bosons multiplets transforming under
bifundamental representations  like $(M_i,{\overline M}_{i+1})$.
However, each  $i^{th}$ such multiplet transforms as
a chiral multiplet under a different  $\cn=1$ subgroup of the $\cn=4$.
Thus, as a whole the theory respects no supersymmetry at all.
However, the spectrum remains Bose-Fermi degenerate up to
one-loop and it is only broken at two loops by diagrams
involving different chiral multiplets of different
$\cn=1$ supersymmetries.
One can obtain such type of field theories from simple
D6-brane configurations on Type IIA string theory
compactified on a six torus $T^6=T^2\times T^2\times T^2$.
The gauge group of D6-branes on $T^6$ respects $\cn=4$
supersymmetry and has a $U(M)$ symmetry.
On the other hand if we wrap three of the D6-brane
coordinates each one around one of the three $T^2$ tori,
in general different stacks of D6-branes will intersect
\cite{bgkl,afiru,bkl,imr}.
At the intersections chiral fermions in bifundamental
representations appear. In addition massless scalars do also
appear for certain intersecting angles. Thus the
above structure of Q-SUSY field theories
is obtained.
In fact we found  the structure of Q-SUSY models
while studying the SUSY properties of the toroidal
intersecting models of ref.\cite{imr}. Recently
ref.\cite{acg} appeared in which
a class of models with n-loop-suppression
of quadratic divergences was presented.
Although in our models there is a two-loop suppression 
of quadratic divergences, the models we find are essentially
different in several respects.

The structure of the present paper is as follows. In the next section  
we describe the idea of quasi-supersymmetric field theories in general
and present a graphic (SUSY quiver) way to describe some of their
properties.
In Section 3 we briefly recall toroidal and orientifold compactifications
of Type II string theory with wrapping D6-branes intersecting at angles.
These  systems provide us with specific string constructions
of Q-SUSY field theories. In Section 4 we describe the conditions under which
those theories give rise to Q-SUSY brane configurations. For certain
complex structure values and brane wrapping numbers the property
of Q-SUSY is obtained. RR tadpole cancellation conditions turn out to
be very constraining and one can see that it is impossible to
get full $\cn=1$ SUSY models, only Q-SUSY models may be obtained from
D6-branes wrapping a six torus. However we will show that one can construct
certain brane models in which a subsector of the theory has $\cn=1$
invariance with other $\cn=0$ subsectors acting as some sort of
``hidden sectors'' of the theory.

Some aspects of the effective field theory of the Q-SUSY
brane configurations are discussed in Section 5. 
In particular
we compute the gauge coupling constants and Fayet-Iliopoulos 
terms from the DBI action and show how they match the
field theory expectations from holomorphicity.
 These results apply as well to
$\cn=1$ supersymmetric configurations like those studied
in \cite{csu}. 
Small variations of the complex structure around the
$\cn=1$ (or Q-SUSY) points give rise to 
such Fayet-Iliopoulos terms. Whereas in the 
$\cn=1$ case the turning on of a FI-term indices
gauge symmetry breaking but not SUSY-breaking, 
in the case of Q-SUSY brane configurations 
the FI-terms may induce in general both 
gauge and local SUSY breaking.

Although, as expected, the NS complex structure
potential in the case of Q-SUSY brane configurations does not vanish 
(leading to NS-tadpoles),
we show that it  has a simplified form compared to generic toroidal 
non-Q-SUSY configurations. In particular we show that such 
scalar potential is linear in the complex structure fields
and that in particular cases the $D=4$ dilaton tadpole vanishes
as a consequence of RR-tadpole cancellations.
 
We present some specific models in Section 6. In particular
we present a model with the spectrum of the SUSY SM, three
quark-lepton generations  and a doubled Higgs sector.
In this realistic example quadratic divergences only
appear at two-loops, providing an specific example of
an stabilized $M_s/M_Z$  hierarchy as mentioned above.
In the present context the SM Higgs mechanism has a
nice geometrical interpretation as a recombination of the
three branes supporting the gauge group $U(2)_L\times U(1)$ 
into a single brane. 
We present some general comments and
conclusions in Section 7.

\section{Quasi-Supersymmetric Models }

A  Q-SUSY  field theory is one in which different
subsectors of the theory respect some $\cn=1$ supersymmetry
but different sectors clash with each other so that the 
complete Lagrangian has no surviving supersymmetry.
The gauge interactions have extended $\cn=4$ (or $\cn=2$) 
supersymmetry whereas the chiral matter fields 
respect only some of the $\cn=1$ subgroups.
Thus, for example, a possible general structure 
for the Lagrangian is as follows
\footnote{We concentrate in this class because it is the one
which appears most easily in explicit D-brane
toroidal models.}
:
\beq
L \ =\ L(\cn=4) \ +\  \sum _i L_i(\cn=1)    \ i=1,2,3,4
\eeq
where $i$ labels  four independent supersymmetries inside $\cn=4$.
Here $L(\cn=4)$ is the $\cn=4$ Lagrangian of a number of group
factors, i.e. $\prod_a U(N_a)$. 
On the other hand each of the
$L_i(\cn=1)$  terms respect a different $\cn=1$ subgroup of the   
full $\cn=4$. 
There will be chiral multiplets
$\Phi _a(\phi_a, \psi_a)$ with
respect to the four different $\cn=1$'s. They will transform
as bifundamentals under the $\prod _{a}U(N_a)$
gauge group
\beq
\Phi_a \ =\ (N_a, {\overline N}_{a+1})
\eeq
or 
\beq
\Phi_a \ =\ (N_a, { N}_{a+1})
\eeq
Each of these chiral multiplets may come in several copies.
Such theories will in general be chiral and one will have
to insure the cancellation of anomalies. In the case
of models obtained from D-brane constructions that will be 
guaranteed by cancellation of RR-tadpoles.
Note that the $a^{th}$ chiral multiplet couples to some $i^{th}$
gaugino but not to the other three $\cn=4$ gauginos. So each
of the boson-fermion multiplets form a chiral multiplet with respect
to a different $\cn=1$ subgroup. It is clear that from the point
of view of one of the $\cn=1$'s the other break explicitly
supersymmetry since its gauginos do not connect the other scalars into
their fermionic partners.
Thus the complete theory will have no supersymmetry unbroken, 
all supersymmetries are broken explicitly.

\EPSFIGURE{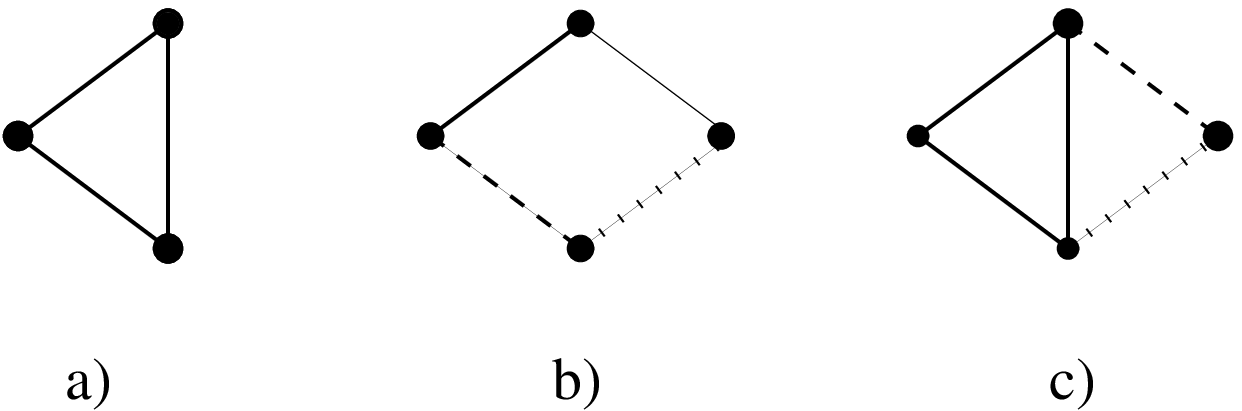, width=5in}
{\label{sefir1}
Some quiver-like graphs
representing SUSY and Q-SUSY field theories:
a) A SUSY theory with three gauge factors; b) A Q-SUSY model
with four group factors and chiral bifundamental fermions filling
SUSY multiplets with respect to four different SUSY's; c)
A Q-SUSY model with four group factors and fermions inside
SUSY multiplets with respect to three different SUSY's.}

One can represent graphically this class of theories
in a quiver-like notation, as in the examples
depicted in figure \ref{sefir1}.
Here the blobs denote $U(N_a)$ factors with $\cn=4$ 
supersymmetry. The links represent the bifundamental
chiral multiplets $\Phi_a$. The different style of the
lines indicate that the boson and fermion in the
multiplet are partners with respect to a different $\cn=1$ SUSY 
inside $\cn=4$. Analogous quiver-like structure  of gauge theories
has arisen in recent years both in the context of string theory
(see e.g.\cite{quivers}) and field theory\cite{decons}.

It is clear from the structure of this class of
theories that, since the gauge
multiplet conserves all four SUSY's, up to one loop
the usual no-renormalization theorems apply for each of
the four $\cn=1$'s independently. Thus quadratic divergences will be
absent up to one-loop. 
The first loop corrections (fig.\ref{twoloops}-a)) to the masses of
the scalars $\phi_a$ 
will appear at two-loop order
\beq
m_a^2 \ \propto [ ( { {\alpha_a}\over {4\pi }})^2 +
 ( { {\alpha_{a+1}}\over {4\pi }})^2 ] \ M^2
\eeq
where $M$ will be some cut-off scale (the string scale in the
D-brane examples discussed below).
As we will see later on, specific D-brane realizations of the
Q-SUSY structure have in addition sectors which are
{\it massive} and respect no supersymmetry at all. These
$\cn=0$ sectors in general decouple but do contribute in
loops to the effective action of the massless fields. 
In particular those massive fields also contribute  
in two-loop order (fig.\ref{twoloops}-b)) to the masses of the
 scalars and at one-loop give masses to
the fermion fields in the $\cn=4$ gauge multiplets 
 (fig.\ref{gauginos})). Adjoint scalars get their masses from diagrams
analogous to those in fig.2-b.
Note that in the absence of these massive $\cn =0$ sectors only the
scalars would get (two-loop) masses but not the gauginos.

\EPSFIGURE{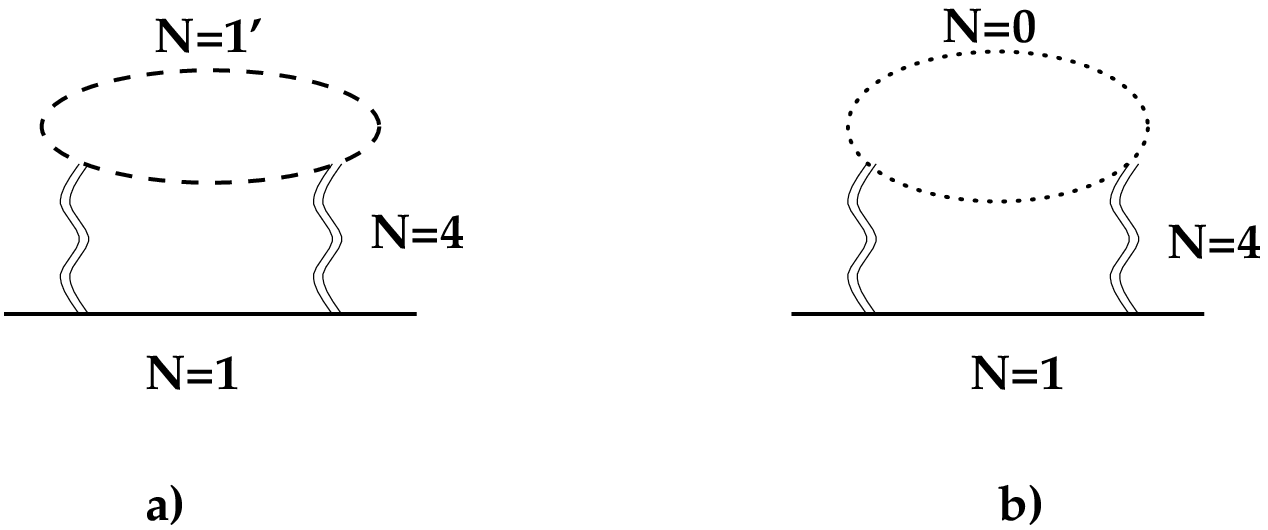, width=4in}
{\label{twoloops} 
 First non-vanishing  loop contributions to the  masses of
scalar fields in a Q-SUSY model: a) Quadratic divergent
contribution present  when the upper loop contains
{\it massless }  fields
respecting different supersymmetries than those of
the  fields below ; b) Contribution coming from
possible  {\it massive } non-supersymmetric states
in the upper loop.}

\EPSFIGURE{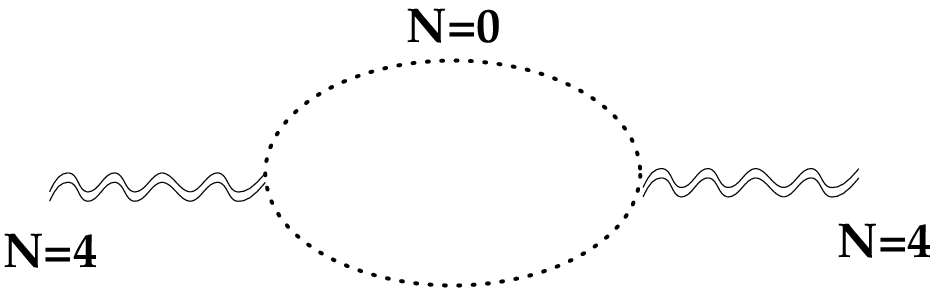, width=3in}
{\label{gauginos} 
One-loop contribution to the masses of gauginos.
 Heavy non-supersymmetric states 
circulate in the loop.}

We thus see that the scalars in this class of theories 
have a two-loop protection against quadratic divergences.
This in general will not be sufficient to solve the
large hierarchy problem between the weak-scale and the Planck
scale
\footnote{See however footnote 17.}. 
However it has been recently realized that the 
scale of new fundamental physics may well be much below,
at scales of order 10-100 TeV. If this is the case we will have
to explain why the weak scale is 2 or 3 orders of magnitude 
below this new scale. In this connection the partial
protection against quadratic divergences of
Q-SUSY models could be of some interest. We will present
later on some semi-realistic D-brane models making use of
this possibility and leave further phenomenological applications 
of this idea to a separate  publication.

The idea of Q-SUSY field theories may be considered independently
of any string theory argument. However it is in the realm of
string theory and brane physics that  it looks 
more natural. In the rest of this paper we are
going to 
present explicit realizations of the idea in terms of Type II
D6-branes wrapping a six-torus. In Section 6 some explicit 
D-brane models are presented.

\section{D6-branes wrapping intersecting cycles in toroidal and orientifold 
compactifications}

Let us describe a simple string setting where the idea of Q-SUSY models
can easily be realized. Consider type IIA string theory compactified
on a six dimensional manifold $\cam$. When constructing a brane
configuration, our building blocks will consist of D6-branes filling
four-dimensional Minkowski space-time and wrapping homology 3-cycles
of $\cam$. A specific configuration of branes will be given by $K$ stacks
of branes, each stack containing $N_a$ coincident D6-branes wrapping the
3-cycle $[\Pi_a] \in H_3(\cam,\IZ)$, ($a = 1,\dots,K$). The gauge group of
such configuration will be given by $\prod_a U(N_a)$. There is a chiral
fermion \cite{bdl} living at each four-dimensional intersection of two   
branes $a$ and $b$, transforming in the bifundamental representation of
$U(N_a) \t U(N_b)$. The intersection number of these two branes,
$I_{ab} \equiv \Pi_a \cdot \Pi_b$ is a topologically invariant integer
whose modulus gives us the multiplicity of such massless fermionic content
and its sign depends on the chirality of such fermions. 

Any consistent configuration has to satisfy some conditions related to the
propagation of RR massless closed string fields on the compact manifold
$\cam$. These are the  RR tadpole cancellation conditions
which can be expressed
in a very
simple way in the context of type IIA D6-branes wrapping 3-cycles. Namely,
the sum of the homology cycles where the branes wrap must add up to zero
\cite{afiru,angel2}
\beq
\sum_a N_a [\Pi_a] = 0.
\label{tadpoles}
\eeq
When considering theories where additional sources for RR charges appear,
such as orientifold compactifications, this topological sum of RR charges
must cancel the  charge induced by the O6-plane. It can be easily
seen that RR tadpoles conditions directly imply the cancellation of
non-abelian $SU(N_a)^3$ anomalies. They also imply, by the
mediation of a Green-Schwarz mechanism, the cancellation of mixed non-abelian
and gravitational anomalies \cite{afiru,imr,csu}.

A particularly simple subfamily of the configurations described above consist
of taking $\cam$ as a factorizable six-torus: $\cam = T^2  \t T^2 \t T^2$.
We can then further simplify our configurations by considering branes wrapping
factorizable 3-cycles, that is, cycles that can be expressed as products of
1-cycles on each $T^2$. This setup and some orbifold and 
orientifold variations have previously been considered in 
\cite{bgkl,afiru,bkl,imr,bklo,csu,inter1}, where explicit
configurations have been constructed
\footnote{For some related constructions involving branes at angles
other than D6-branes see \cite{afiru,inter2,cim3}. For some work
concerning phenomenological aspects of these constructions see
\cite{afiru2,imr,bklo,pheno,cim2}.}. In this case the homology 3-cycle
$\Pi_a$ can be expressed as
\beq
[\Pi_a] \equiv [(n_a^1,m_a^1),(n_a^2,m_a^2),(n_a^3,m_a^3)],
\label{factorizable}
\eeq
where $n_a^i, m_a^i$ being integers describing
wrapping numbers around the tori cycles. The intersection number
takes
the simple form
\beq
I_{ab} = \prod_{i=1}^3 \left(n_a^i m_b^i - m_a^i n_b^i \right).
\label{intersection}
\eeq
whose modulus gives us the number of chiral fermions at the intersection.
An interesting variation of toroidal compactifications consist of modding
out our six-torus by the $\IZ_2$ orientifold group action $\{1, \Om\car\}$,
 where $\Om$ is the world-sheet parity and $\car$ is a reflection on the
vertical coordinates $x^5,x^7,x^9$. This system naturally arises by   
considering type I string theory compactified on $T^6$, where D9-branes   
with fluxes appear, and performing a T-duality on these vertical
coordinates \cite{bdl,bgkl,bkl,torons}
\footnote{See also \cite{Bachas,fluxes,bfield}}.
These orientifold compactifications have several
new features compared to the toroidal ones. First, an O6-plane appears in the
compactification, wrapping a 3-cycle $[\Pi_{ori}]$. Second, to each stack
of branes $a$ we must add its image under the orientifold group generator
$\Om\car$. Thus, to each sector $D6_a$ we must add a mirror sector
$\Om \car D6_a$ or $D6_{a^*}$. When dealing with square tori, tadpoles read
\beq
\sum_a N_a \left( [\Pi_a] + [\Pi_{a^*}] \right) = 32 \ [\Pi_{ori}].
\label{tadpoles2}
\eeq
and it can easily be seen by performing  
the above mentioned T-duality  that we are left with an O6-plane on the
3-cycle $[(1,0),(1,0),(1,0)]$. If a brane is wrapping the homology cycle 
(\ref{factorizable}), then its mirror brane is given by
\beq
[\Pi_{a^*}] \equiv [(n_a^1,-m_a^1),(n_a^2,-m_a^2),(n_a^3,-m_a^3)].
\label{mirror}
\eeq
In a configuration composed by factorizable branes, then, the tadpole
conditions read \cite{bgkl}
\begin{eqnarray}
\sum_a N_a n_a^1 n_a^2 n_a^3 = 16  \nonumber\\
\sum_a N_a m_a^1 m_a^2 n_a^3 = 0 \nonumber\\
\sum_a N_a m_a^1 n_a^2 m_a^3 = 0 \nonumber\\
\sum_a N_a n_a^1 m_a^2 m_a^3 = 0.
\label{tadpoles3}
\end{eqnarray}
Notice that only four conditions appear in (\ref{tadpoles3}), in contrast
with the eight conditions appearing in the toroidal case (see \cite{afiru}).
This signals that only four four-dimensional RR fields are relevant for
tadpoles under this orientifold compactification. Similarly, some of the
NSNS fields that describe the geometry of the $T^6$ are no longer
dynamical. This is because not any complex structure is well defined under
the identification $\Om \car$, which only allows square tori and a special
kind of tilted tori (see figure \ref{bflux}). In the T-dual picture,
this translates into a
B-field non-invariant under $\Om$, which only allows discrete values
$b = 0, \oh$ \cite{bfield,bkl}. In order to easily
describe these configurations with non-vanishing $b$-flux, we define 
effective wrapping numbers, which can now take semi-integer values
\beq
(n_a^i,m_a^i)_{{\rm eff}} \equiv (n_a^i,m_a^i) + b^{(i)} (0,n_a^i),
\label{fwrapping}
\eeq
where $b^{(i)}$ stands for the value of $b$ in the $i^{th}$ $T^2$.
Expressing our factorizable D6-branes in terms of these fractional
wrapping numbers makes (\ref{mirror}) and (\ref{tadpoles3}) still
valid in the presence of non-trivial $b$-flux. Notice, however, that
the O6-plane will now lie in the 3-cycle
$[({1 \over \beta^1},0)_{{\rm eff}}, ({1 \over \beta^2},0)_{{\rm eff}},
({1 \over \beta^3},0)_{{\rm eff}}]$, where $\beta^i = 1-b^{(i)}$.
\footnote{From now on we will suppress the subindex 'eff', considering
always fractional wrapping numbers.}

\EPSFIGURE{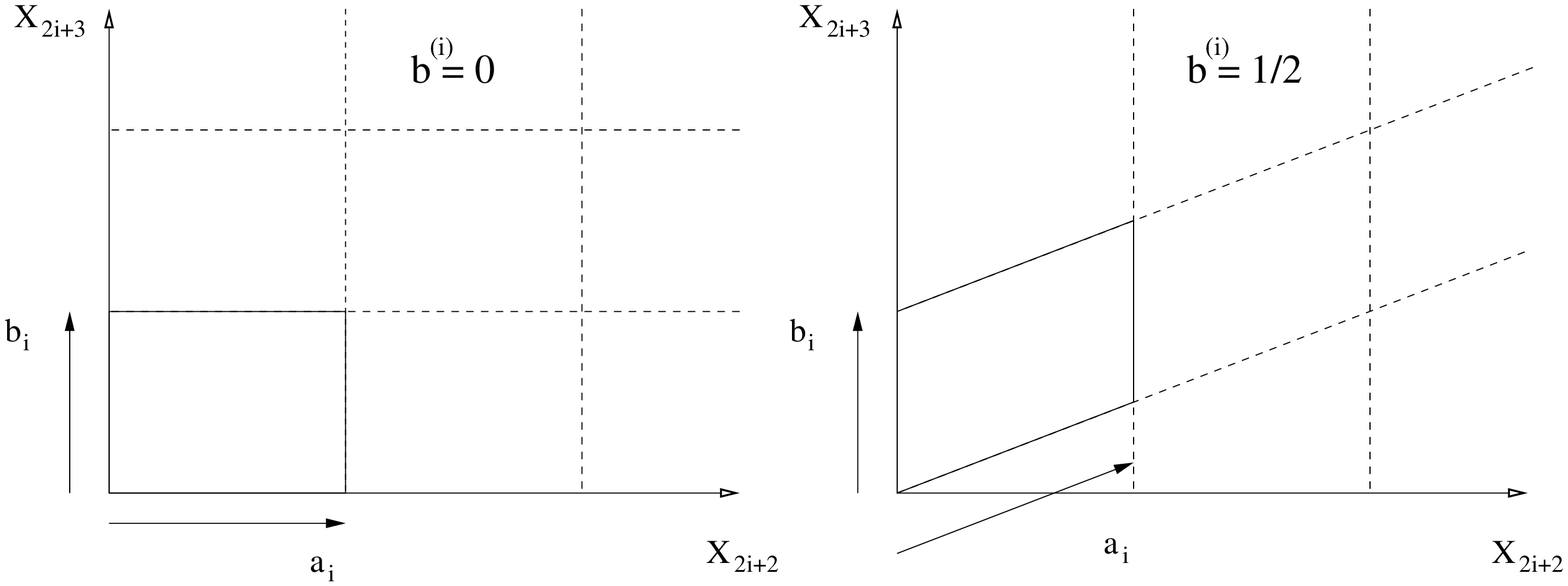, width=6in}
{\label{bflux} 
Allowed toroidal lattices in orientifold models.}

Let us finally point out that considering orientifold models introduces
new sectors arising from open strings stretching between branes. To the
previous $D6_a D6_a$ sector, giving rise to unitary gauge group $U(N_a)$,
\footnote{In orientifold models, also $SO(N)$ and $USp(N)$ gauge groups
may appear if a brane $a$ is its own image under the orientifold action, 
but this will not usually be the case in our constructions.} 
and the $D6_a D6_b$ sector,
where massless fermions in the bifundamental $(N_a, \ov N_b)$ live,
we must add the $D6_a D6_{b^*}$ sector, with massless fermions in
the  $(N_a, N_b)$ representation, and the $D6_a D6_{a^*}$, where  
'exotic matter' transforming in the symmetric and antisymmetric
representation of $U(N_a)$ may appear (for more details see
\cite{bgkl,imr}).

\section{Systems of branes preserving a Supersymmetry}

Let us consider IIA D6-branes wrapping homology 3-cycles on a six dimensional 
manifold $\cam$. If we consider an isolated brane in a given homology class 
$[\Pi_a] \in H_3(\cam,\IZ)$, this brane will tend to minimize its tension by 
wrapping the submanifold of minimum length inside this homology class
\footnote{This is strictly true only when no $B$ or $F$ fluxes are turned 
on, so the minimization of the DBI action is equivalent to the minimization
 of the volume.}. A 
particular class of volume-minimizing submanifolds are  Special 
Lagrangian submanifolds (SLAGs) which appear naturally as calibrated objects
in the geometry of Calabi-Yau compactifications. These manifolds have been 
discussed recently in the literature \cite{Joyce,bbh,Denef,CY}. 
The precise form, and even the existence of these SLAGs, 
depends on the specific point of the moduli space of complex structures 
of $\cam$ we are sitting on. A brane wrapping a Special Lagrangian
submanifold will preserve a supersymmetry, thus studying SLAGs yields an 
interesting family of BPS states in a particular compactification.
A full configuration of D6-branes may, however, be composed of 
branes that preserve different supersymmetries, thus leading to a 
non-supersymmetric system. 


\subsection{System of two branes}

In general, we can detect whether two branes $D6_a$ and $D6_b$ preserve 
a common supersymmetry by looking at the spectrum living at the $D6_aD6_b$ 
sector. 
A particularly simple setting consist of D6-branes wrapping factorizable 
cycles of a $T^2 \times T^2 \times T^2$. By factorizable we mean branes 
wrapping a 1-cycle on each $T^2$, so we are actually considering a sublattice 
of the whole $H_3(T^6,\IZ)$. The SLAGs corresponding to these factorizable 
cycles are the product of three straight lines wrapping the corresponding 
1-cycle on each torus, thus yielding a flat 3-submanifold of $T^6$. Each of 
these branes preserves the maximum number of supersymmetries, which is 
$\cn = 4$.

We may consider a system of two branes: $D6_a$ and $D6_b$, both wrapping a 
factorizable cycle on our six-dimensional torus. As both branes make an 
angle $\vt_{ab}^i$ on the $i^{th}$ torus, we can describe the spectrum in 
the $D6_aD6_b$ sector by introducing a four-dimensional twist vector $v_\vt$ 
\cite{afiru}
\beq
v_\vt = (\vt_{ab}^1,\vt_{ab}^2,\vt_{ab}^3,0),
\label{twist}
\eeq
where the fourth component describes the non-compact complex dimension in 
light-cone gauge. Here we are considering $\vt_{ab}^i$ in units of $\pi$, 
so that $-1 \leq \vt_{ab}^i \leq 1$. The states living in this sector can 
be described by the sum of vectors $r + v_\vt$, where $r \in (\IZ + \nu)^4$. 
The GSO projected states are those that satisfy $\sum_i r^i = odd$, both 
for the R sector ($\nu = 0$) and the NS sector ($\nu = \oh$). The mass of 
these states is given by \cite{arfaei,afiru}
\beq
\a' M_{ab}^2 = {Y^2 \over 4\pi\a^\prime} + N_{bos}(\vt) + {(r + v_\vt)^2 
\over 2} -\oh + E_{ab},
\label{mass}
\eeq
where $N_{bos}(\vt)$ is the contribution from the bosonic oscillators and  
$E_{ab}$ is the vacuum energy:
\beq
E_{ab} = \sum_{i=1}^3 \oh |\vt^i| (1 - |\vt^i|)
\label{vacuum}
\eeq

Given a generic twist vector $v_\vt$ with non-trivial angles $\vt_{ab}^i$, 
we will always find a massless fermion in the R sector, described by the 
vector 
\beq
r_R = \oh \left(-{s(\vt_{ab}^1)}, -{s(\vt_{ab}^2)}, -{s(\vt_{ab}^3)},
{\prod_{i=1}^3 s(\vt_{ab}^i)}\right),
\label{fermion}
\eeq
where $s(\vt_{ab}^i) \equiv sign(\vt_{ab}^i)$. This massless fermion will 
live at the submanifold where both branes intersect, which is generically 
one or several points in the compact space (times the full four-dimensional
Minkowski space). The fourth component of $r$, 
which describes the four-dimensional Lorentz quantum numbers of the state, 
is fixed by GSO projection, and it is easy to see that its sign agrees with 
that of the intersection number $I_{ab}$.

Looking at the lightest states coming from the NS sector we find four 
different scalars with masses given by \cite{afiru}:
{\small \beqa
\begin{array}{cc}
{\rm \bf State} \ (r_{NS}) \quad & \quad {\bf Mass^2} \\
(-s(\vt^1),0,0,0) & \alpha' M^2 =
\frac 12(-|\vartheta^1|+|\vartheta^2|+|\vartheta^3|) \\
(0,-s(\vt^2),0,0) & \alpha' M^2 =
\frac 12(|\vartheta^1|-|\vartheta^2|+|\vartheta^3|) \\
(0,0,-s(\vt^3),0) & \alpha' M^2 =
\frac 12(|\vartheta^1|+|\vartheta^2|-|\vartheta^3|) \\
(-s(\vt^1),-s(\vt^2),-s(\vt^3),0) & \alpha' M^2
= 1-\frac 12(|\vartheta^1|+|\vartheta^2|+|\vartheta^3|)
\label{scalars}
\end{array}
\eeqa}

Notice that these four scalars may be massive, massless or even tachyonic, 
depending on the relative angles both branes make. In general, when having 
one of these four scalars massless, the sector $D6_aD6_b$ will present a 
degeneracy in mass between bosonic and fermionic states, all the spectrum of 
particles arranging themselves into D=4, $\cn = 1$ supermultiplets. This 
signals that, for this particular choice of complex structure, our combined 
system of D6-branes preserves a common SUSY. It may also happen that two 
(or even four) of the scalars become massless, yielding a D=4 $\cn = 2$ 
($\cn = 4$) spectrum \cite{imr,torons}.

The supersymmetry preserved for such system can be characterized by a vector 
$\tilde r \in (\IZ +\oh)^4$ with opposite GSO projection, defined by $\tilde r 
\equiv r_{NS} - r_R$, where $r_{NS}$ corresponds to a massless scalar. 
Indeed, we can associate to $\tilde r$ a SUSY generator
$Q_{\tilde r}$, 
that takes us from a massless fermionic state to a massless bosonic state:
\beq
Q_{\tilde r} |r + v_\vt>_R = |r+\tilde r + v_\vt>_{NS}.
\label{generator}
\eeq

This connection between bosonic and fermionic states will not only hold at 
the massless level, but will also be true for any massive supermultiplet in 
the $D6_aD6_b$ sector.

\subsection{System of several branes}

When facing the problem of building a chiral theory arising from the low
 energy description of a fully-fledged compactification of branes 
intersecting at angles, usually more than two stacks of branes have to 
be considered. This is mainly due to the fact that, in order to have a 
consistent anomaly-free 4D theory, some constraints have to be satisfied,
 namely the RR tadpole cancellation conditions
. As we have 
seen in Section 3, RR tadpoles cancellation translates into a topological
 restriction on the sum of the homology classes where the D6-branes wrap.
 Briefly stated, this sum has to be equal to the homology class of the 
O6-plane, if present.

When studying supersymmetric compactifications, however, we realize that 
the condition for a pair of factorizable branes to preserve a SUSY depends
 on the angles they make on each of the tori. Supersymmetry, then, turns 
out to be a geometrical question, rather than a topological one. By varying 
the complex structure of the manifold where the D6-branes wrap, we can go
 from a SUSY configuration to a non-SUSY one. A natural question is whether
 we can go the other way round, that is, if any generic compactification 
allows for a nontrivial chiral SUSY model by suitably changing the complex 
structures.

This question can be made more precise. It is well known that two 
factorizable D6-branes at angles preserve a supersymmetry if they are
 related by a rotation which belongs to $SU(3)$ \cite{bdl,polchi2}. 
Now, this $SU(3)$  rotation can be embedded in several ways into the 
tangent space rotation group $SO(6) \simeq SU(4)$. 
Let us describe the relative position between two factorizable
branes $a$ and $b$ by a rotation matrix acting on the compact complex 
coordinates $z^i = x^{2i+2} + i x^{2i+3}$ that parametrize each of
the tori $T^2_i$ ($i = 1,2,3$), as done in \cite{polchi2}.
\beqa
R_{ab}: \left(
\begin{array}{c}
z^1 \\ z^2 \\ z^3
\end{array}
\right) \mapsto
\left(
\begin{array}{ccc}
e^{i\pi\vt^1} & 0 & 0 \\
0 & e^{i\pi\vt^2} & 0 \\
0 & 0 & e^{i\pi\vt^3}
\end{array}
\right)
\cdot
\left(
\begin{array}{c}
z^1 \\ z^2 \\ z^3
\end{array}
\right)
\label{rotation}
\eeqa

In general, $R$ belongs to a $U(3)$ subgroup of $SO(6)$ that
preserves the complex structure of $T^6$. Since we have chosen
it to relate two factorizable branes which are 1-cycles on each
$T^2$, it is also diagonal. The conditions for preserving a 
SUSY can be read from the previous discussion,
in terms of the masses (\ref{scalars}). This translates into a restriction 
on $R$, which is $\oh \sum_i \vt^i \eps_i \in \IZ$ for some phases 
$\eps_i=\pm 1$.\footnote{More strictly speaking, this restriction comes from 
imposing $R_{ab}^2 Q_{\tilde r} = Q_{\tilde r}$ for some $\tilde r$, 
where $Q_{\tilde r}$ is one of the ten-dimensional spinor 
states appearing in Type IIA string theory. See \cite{jabbari,bgkl}.}
Notice that this implies $R \in SU(3)$ (or some isomorphic
embedding of this group into $SO(6)$).

This fact can easily be applied to a full configuration of branes, where 
the same SUSY will be preserved by any of the branes if they are related to
 each other by rotations of the same $SU(3)$. So, in order to see if we have
 a supersymmetric configuration of factorizable branes, we only have to 
compute the rotations $R_{ab}$ relating each pair of D6-branes $a$ and $b$
 and check that they all belong to the same $SU(3)$ subgroup. This will 
mean, in turn, that the vector $\tilde r$ at each intersection is the same.

\subsection{SUSY and Q-SUSY orientifold systems}

There is a simple way to show that chiral supersymmetric configurations are 
impossible to obtain in toroidal models of D6-branes considered in 
\cite{afiru}. Indeed, it is very easy to construct RR tadpole-free 
configurations of this kind of compactifications. However, having 
supersymmetric configurations would automatically imply the cancellation 
of NSNS tadpoles, which are proportional to the sum of tensions of the 
branes. As mentioned in \cite{bklo}, this can easily be seen by the
dimensional reduction of the whole Dirac-Born-Infeld action for 
every D6-brane. 
\beqa
{\cal S}_{DBI} & = & - T_6 \sum_a \int_{D6_a} d^{p+1}\xi \
e^{-\phi} \sqrt{det(G + \cf_a)}
\nonumber\\
& = & - \int_{\cam_4} d^4x \ {T_6 \over \lambda} \
\sum_a\A l_a \A.
\label{NSpotential}
\eeqa
Here, $T_6$ is the tension of the D6-branes  from the 
general formula $T_p = (2\pi)^{-p} \a^{\prime \ -\frac{p+1}{2}}$
\cite{polchi2}. $\lambda$ is the string coupling and $\A l_a \A$ 
stands for the
volume of the 3-manifold where the brane lies inside the compact space $T^6$. 
Since this quantity only vanishes if the sum of tensions is null, that is if 
no branes are present, we deduce that the only way to have a supersymmetric
 configuration is just having Type IIA string theory compactified on $T^6$,
with no branes at all.
 In order to avoid this result, we can perform some orientifolding
 on our configuration. This will introduce some negative contribution to the
 NS potential, coming from the negative tension of the O6-plane
\beq
V = {T_6 \over \lambda } 
\left( \sum_a \A l_a \A - \A l_{ori} \A\right).
\label{NSpotential2}
\eeq

This allows us to have some brane content when canceling both RR and NSNS 
tadpoles. Notice, however, that this may not be good enough. If the
 O6-plane lies in just one factorizable 3-cycle of $T^6$, as in orientifold
 models considered in \cite{bgkl,bkl,imr}, then the supersymmetric 
configuration
 will only be attained when every D6-brane lies on top of the orientifold 
(or parallel to it), yielding just a T-dual version of Type I string theory
 compactified on a torus. In order to attain a chiral supersymmetric 
configuration we should have, then, several O6-planes wrapping different
 cycles \footnote{Or, equivalently, a O6-plane wrapping a non-factorizable
 cycle.}. In \cite{csu} a family of $\IZ_2 \t \IZ_2$ orientifold models was
 presented, where the O6-plane wrapping a definite factorizable 3-cycle had
 to be completed by its images under the orbifold $\IZ_2 \t \IZ_2$ group 
action. 

Adding some negative tension to (\ref{NSpotential2}) is not  the 
only effect of considering an orientifold compactification. When constructing 
a toroidal model, a factorizable brane $a$ by itself, being a $\oh$BPS state 
will preserve $\cn = 4$ supersymmetry. In an orbifold or orientifold model, 
however, this brane has to share some supersymmetry with its images under the
 orientifold group action. Stated differently, the brane $a$ has to share 
some supersymmetry with the O6-plane, that is, it should be related by a 
$SU(3)$ rotation (\ref{rotation}) with the O6-plane. If the O6-plane is 
wrapping a factorizable cycle, then it preserves, so to speak, $\cn = 4$,
 and it can share a supersymmetry with a brane in many different ways.
In the models presented in \cite{csu}, however, the $\IZ_2 \t \IZ_2$
twist was already embedded as a discrete group of a definite $SU(3)$
rotation group. This has two consequences. First, the ambient space 
will only preserve $\cn = 1$ (as in usual Calabi-Yau orientifold theories).
Second, if a brane is to preserve a SUSY with its image under the 
orientifold group, then it should be related to the O6-plane(s) by a
rotation inside this same $SU(3)$. In the explicit constructions presented
in \cite{csu} this condition translated into $\vt^1 + \vt^2 + \vt^3 \in \IZ$.
Indeed, the branes considered in model building had the following 
twist vectors:
\beq
\begin{array}{lr}
a \ {\rm branes}: & v_a = (0,\vt_a,-\vt_a) \\
b \ {\rm branes}: & v_b = (\vt_b,0,-\vt_b) \\
c \ {\rm branes}: & v_c = (\vt_c,-\vt_c,0) 
\end{array}
\label{vectors}
\eeq

The supersymmetry preserved by any intersection of these branes with the 
O6-plane or any of its images under  $\IZ_2 \t \IZ_2$ is given by the vector 
$\tilde r = \pm\oh (+,+,+,+)$. By transitivity, any intersection of branes 
also preserves this same supersymmetry. Notice that $\tilde r$ is the 
only spinor invariant under the $\IZ_2 \t \IZ_2$ orbifold group, so the 
only candidate SUSY to be preserved by a pair of branes. This argument 
seems general for any orientifold compactification with the O6-plane 
wrapping a non-factorizable cycle.

The setup presented in \cite{csu} allowed for the construction 
of an interesting class of chiral supersymmetric models of branes at
 angles, as was explicitly shown. However, the introduction of the 
orbifold twists is not harmless from the model building point of view 
and the massless chiral spectrum tends to yield particles
with exotic quantum numbers. In this paper we will
concentrate in the simpler case of toroidal orientifolds 
without orbifold twists, although some of our results,
particularly those related to the effective Lagrangian 
in Section 5 will remain valid in the orbifold cases.

When considering an orientifold compactification where the O6-plane lies 
in a factorizable cycle we can consider a different possibility when 
constructing a ``supersymmetric'' model, which might be of 
phenomenological interest. Instead 
of asking to our branes to share the same SUSY $\tilde r$ with the O6-plane, 
we can  relax this condition and ask them to preserve at least one SUSY with
 it, but not necessarily the same one for each brane. In this setup, a pair
 of branes intersecting at a point may, as well, share a SUSY, thus having 
a boson-fermion degenerate mass spectrum. If this happens for every pair of
 intersecting branes, then, for each massless fermion living at the 
intersections we will have a massless boson superpartner. Since each 
intersection preserves a different SUSY, we will effectively have $\cn = 0$
 when considering the full field theory at low energy. However, locally 
(at each intersection) particles will arrange as SUSY multiplets. 
This is an explicit D-brane realization of the Q-SUSY idea introduced 
in Section 2.

Let us give an example realizing this idea. We will consider 
Type IIA D6-branes wrapping factorizable cycles of a torus 
orientifolded by $\Om \car$. In this particular setting, branes are
allowed to preserve $\cn = 2$ SUSY with the orientifold and their mirror 
images. We will consider all possible types of such ``$\cn = 2$ branes''
in order to describe the different Q-SUSY structures they may lead to.
A quite general configuration and its possibilities is illustrated
by the brane content shown in table \ref{qmodel1}.
\TABLE{\renewcommand{\arraystretch}{1.4}
\begin{tabular}{|c||c|c|c||c|}
\hline 
 $N_i$ & $(n_i^1,m_i^1)$ & $(n_i^2,m_i^2)$ & $(n_i^3,m_i^3)$ & $v_i$ \\
\hline
\hline $N_{a_1}$ & $(1,0)$  &  $(1,m_a^2)$ &
 $(1,m_a^3)$ & $(0,\vt_a^2,\vt_a^3)$ \\
\hline $N_{a_2}$ & $(1,0)$  &  $(1,m_a^2)$ &
 $(1,-m_a^3)$ & $(0,\vt_a^2,-\vt_a^3)$ \\
\hline $N_{b_1}$ & $(1,m_b^1)$ & $ (1,0)$  &
$(1,m_b^3)$ & $(\vt_b^1,0,\vt_b^3)$ \\
\hline $N_{b_2}$ & $(1,-m_b^1)$ & $ (1,0)$  &
$(1,m_b^3)$ & $(-\vt_b^1,0,\vt_b^3)$ \\
\hline $N_{c_1}$ & $(1,m_c^1)$ & $(1,m_c^2)$  &
$(1,0)$ & $(\vt_c^1,\vt_c^2,0)$ \\
\hline $N_{c_2}$ & $(1,m_c^1)$ & $(1,-m_c^2)$  &
$(1,0)$ & $(\vt_c^1,-\vt_c^2,0)$ \\
\hline 
\end{tabular}
\label{qmodel1}
\caption{\small Example of  D6-brane wrapping numbers giving rise to
a ``$\cn = 2$'' Q-SUSY model. We are supposing $m_i^j \geq 0$, $i = a,b,c$, 
$j = 1,2,3$.}}

The six stacks of branes will give rise to a $U(N_{a_1}) \t U(N_{a_2}) \t 
U(N_{b_1}) \t U(N_{b_2}) \t U(N_{c_1}) \t U(N_{c_2})$ gauge group. 
In order to obtain a non-anomalous 4D theory tadpole cancellations should 
be fulfilled, imposing $N_{i_1} = N_{i_2} = N_i$, $i = a,b,c$.
In order to cancel 
the orientifold charge ($nnn$ tadpole) an additional stack of $N_h$ branes
 parallel to the orientifold may be added. Having wrapping numbers 
$(1,0)(1,0)(1,0)$, it does not intersect with any other brane, so it can be
 considered as a hidden sector of the theory. The last tadpole condition 
reads then
\beq
2N_a + 2N_b + 2N_c + N_h = 16.
\label{nnn}
\eeq

In the last column of table \ref{qmodel1} we introduced the twist vector 
of each stack 
of branes, where $\vt_i^j = tg^{-1} \left(m_a^j R_2^{(j)} /
R_1^{(j)} \right)$, $i = a,b,c$, $j = 1,2,3$.
 Is easy to see that each brane preserves $\cn = 2$ with the O6-plane if and
 only if $\vt_a^2 = \vt_a^3 = \vt_a$, etc$\dots$ Notice that this is only 
possible if $\frac{m_c^2}{m_c^1}=\frac{m_a^2}{m_a^3}\frac{m_b^3}{m_b^1}$. 
This constraint is not present in models where one type of branes 
does not appear (for instance, see the 
models with a square quiver in Section 6 and the phenomenological models 
from \cite{cim2}). 

If we index the possible vectors $\tilde r$ describing a SUSY as
\beq
\begin{array}{c}
\tilde r_1 = \pm \oh (-++-) \nonumber \\
\tilde r_2 = \pm \oh (+-+-) \nonumber\\
\tilde r_3 = \pm \oh (++--) \nonumber\\
\tilde r_4 = \pm \oh (----) 
\end{array}
\label{r}
\eeq
then we can represent the different SUSY's shared by the branes with the 
orientifold plane in table \ref{susys}.
\TABLE{\renewcommand{\arraystretch}{1.1}
\begin{tabular}{|c|c|c|}
\hline
 Brane & Twist vector & SUSY preserved \\
\hline
\hline $a_1$, ($a_1^*$) & $\pm(0,\vt_a,\vt_a)$ & 
$\tilde r_2$, $\tilde r_3$ \\
\hline $a_2$, ($a_2^*$) & $\pm(0,\vt_a,-\vt_a)$ & 
$\tilde r_1$, $\tilde r_4$ \\
\hline $b_1$, ($b_1^*$) & $\pm(\vt_b,0,\vt_b)$ & 
$\tilde r_1$, $\tilde r_3$ \\
\hline $b_2$, ($b_2^*$) & $\pm(-\vt_b,0,\vt_b)$ 
& $\tilde r_2$, $\tilde r_4$ \\
\hline $c_1$, ($c_1^*$) & $\pm(\vt_c,\vt_c,0)$ & 
$\tilde r_2$, $\tilde r_1$  \\
\hline $c_2$, ($c_2^*$) & $\pm(\vt_c,-\vt_c,0)$ & 
$\tilde r_3$, $\tilde r_4$  \\
\hline 
\end{tabular}
\label{susys}
\caption{\small Supersymmetry preserved by each brane with the O6-plane.}}

It is now easy to guess which supersymmetries will be preserved at each 
intersection. In general, an 
intersection $ij$ will preserve the common supersymmetries that 
branes $i$ and $j$ preserve separately with the orientifold. 
This will imply that sectors with non-vanishing intersection number 
(chiral sectors) preserve one of the four SUSY's in (\ref{r}), being six
 such intersections per supersymmetry.
 Sectors with vanishing
intersection number (generally massive)  
contain a $\cn = 2$ subsector, coming from 
brane-brane$^*$ spectrum, and a $\cn = 0$ subsector, coming from 
branes of the same group. This general ``$\cn = 2$'' Q-SUSY structure 
can be expressed in a quiver-like manner as illustrated in figure
\ref{hexagon}. Notice that nodes in this hexagonal diagram 
represent gauge sectors with $\cn = 2$
\footnote{Note that although the combined system of a brane and its 
mirror will respect only $\cn = 2$, as long as they do not overlap
the massless sector will fill $\cn =4$ representations.} 
, and links represent chiral 
sectors where $\cn = 1$ matter multiplets live. When there is no link 
between two nodes this signals a $\cn = 0$ sector, generically massive.
All this Q-SUSY structure concerns the open string sector of the theory,
while it  is embedded in the $\cn = 4$ supersymmetry preserved by the 
closed string sector living in the bulk.

\EPSFIGURE{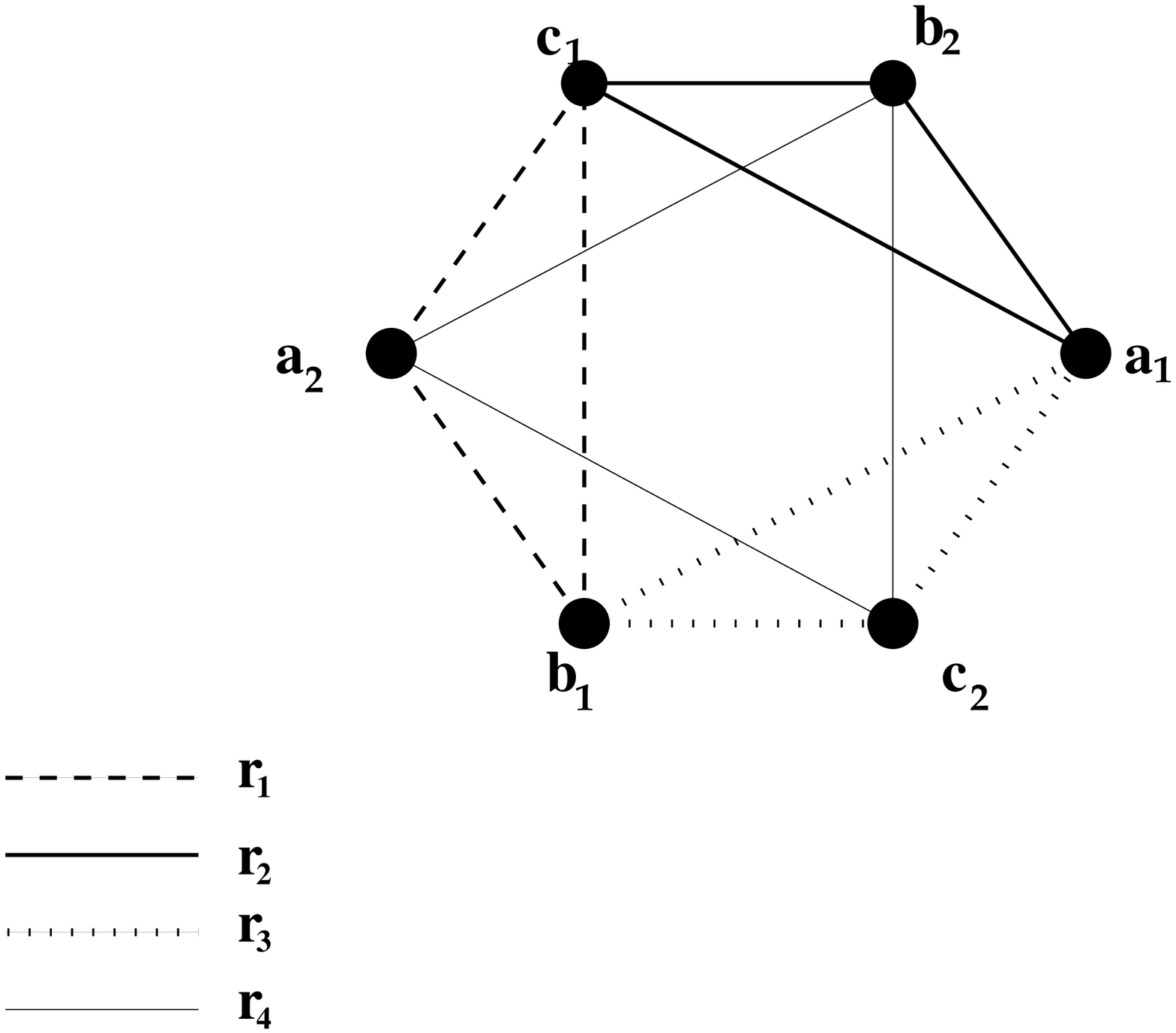, width=3in}
{\label{hexagon}
Quiver diagram corresponding to the 
general class of Q-SUSY models discussed
in the text. The dots represent the six different
types of branes (plus mirrors)  in the models whereas the links represent the 
chiral intersections of those branes. There are four types of links
corresponding to the different SUSY's the corresponding 
intersection preserves.}

Different examples of Q-SUSY brane configurations may be obtained by deleting
some of the nodes, as we will show in specific examples in Section 6.
A comment on superpotentials and Yukawa couplings in the 
general D-brane schemes from the quiver in fig.\ref{hexagon} 
is in order. In explicit D6-brane constructions of Q-SUSY models
some world-sheet instanton effects can communicate sectors
with different $\cn =1$ SUSY's and yield explicit 
SUSY breaking even at the tree level. Indeed, for each triangle 
in fig.\ref{hexagon} with sides of {\it the same type }(hence with chiral
matter respecting the same SUSY) one may have in general
trilinear superpotential couplings respecting the corresponding
$\cn =1$. They come from a disk worldsheet with
 three intersecting branes at the boundary respecting the same $\cn =1$.
However for each triangle in fig.\ref{hexagon} of {\it different type}
there will in general SUSY violating Yukawa couplings, since 
they will couple chiral multiplets respecting 
different supersymmetries. 
They will come from  disk worldsheets  with 
 three intersecting branes at the boundary respecting 
{\it different } $\cn =1$. Note however that the actual
presence and/or relevance of those SUSY-breaking Yukawa 
couplings will be model dependent. In some 
SUSY-quivers like the square quiver in Section 6 
such Yukawa couplings or superpotentials are absent 
(there are no subtriangles in the quiver). In other 
configurations, like the SUSY standard model of
Section 6 only SUSY superpotentials appear.
More generally, in any model the size of some 
SUSY-breaking Yukawa couplings may be exponentially
suppressed if the compact volume is large \cite{afiru2}.
 Thus, for example, in a realistic SM construction it is enough 
if the top and bottom quark Yukawa couplings are supersymmetric, 
since the other Yukawa couplings are very small and would not affect the 
stability of the Higgs mass in a sizable manner.

\section{Effective field theory: Gauge Coupling Constants
and Fayet-Iliopoulos terms}

We would like now to address some aspects of the
effective low-energy field theory at the intersecting
branes. Since we will study the local physics at the 
intersections our results will apply both
to theories with full $\cn =1$ SUSY as well as to 
Q-SUSY theories. We will discuss in turn the 
gauge coupling constants and the Fayet-Iliopoulos terms.
We will also discuss the structure of the NS tadpoles in
Q-SUSY brane configurations.

\subsection{Coupling constants and gauge kinetic functions}

 The gauge coupling constants
 may be computed from the DBI action but also, if
a $\cn =1$ SUSY is preserved at the intersection, from the
real part of a gauge kinetic function. The form of the latter may be
obtained from holomorphicity once we know the imaginary part,
which may be obtained from the known RR-couplings to
gauge fields. We will show here  
how those two independent computations
agree.

When looking at the low energy theory living on the world-volume of a single 
D6-brane we detect, from a four-dimensional point of view, a $U(1)$ SYM 
theory whose gauge coupling constant is controlled by the tension of the 
brane on the compact dimensions. That is, by the length of the 3-cycle the 
brane wraps on the compact manifold $\cam$.
\beq
{1 \over g_i^2} = {M_s^3 \over (2\pi)^4 \lambda} \Arrowvert l_i \Arrowvert,
\label{coupling}
\eeq
where $M_s = \a^{\prime -\oh}$ is the string scale, $\lambda$ is 
the string coupling, and  $\Arrowvert l_i \Arrowvert$ is the 
3-volume of the 3-cycle the brane is wrapping
\footnote{Notice that formula (\ref{coupling}) is the 
correct expression for the $SU(N_a)$ subgroup of $U(N_a)$,
with its generators in the fundamental
representation normalized to unity.  
When computing FI-terms we will be dealing with a $U(1)_a$ subgroup,
whose generator will be taken to be Id$_{N_a}$. Both coupling
constants are then related by $g^2_{(SU(N_a))} = N_a g^2_{(U(1)_a)}$.}.

In a supersymmetric field theory, though, the information concerning the 
gauge coupling constant should be encoded in the gauge kinetic function 
$f_{ab}$ of the theory. This function enters on the supersymmetric 
lagrangian as 
\beq
{\cal L}_g = - {1 \over 4}{\rm Re} f_{ab} \ F_{a\ \mu\nu} F_b^{\mu\nu} + 
{i \over 4}{\rm Im} f_{ab} \ F{a\ \mu\nu} \tilde F_b^{\mu\nu},
\label{lagrangian}
\eeq
from what we deduce 
\beq
{\rm Re} f_{aa} = {1 \over g_{a}^2} = 
{M_s^3 \over (2\pi)^4 \lambda} \Arrowvert l_a \Arrowvert, 
\label{realf}
\eeq

An important property of the gauge kinetic function is its holomorphicity. 
This means that, in a supersymmetric configuration, we should be able to 
express $f$ as an holomorphic function on complex fields. What is more, we 
know what the real part of this function should look like, and it is also 
possible to compute the imaginary part by looking at the world-volume 
couplings of the form
\beq
\int_{M_4} \Phi_i F_a \wedge F_a,
\label{FFcouplings}
\eeq
that should arise in the dimensionally reduced effective theory. Here,
$\Phi_i$ is a four-dimensional dimensionless scalar field. As an 
illustration of all this let us consider the orientifold case  
\cite{bgkl,bkl,imr}. We will consider the T-dual theory to
D6-branes at angles, which is D9-branes (Type I theory) with magnetic 
fluxes. In this theory we have two ten-dimensional RR
fields, $C_2$ and $C_6$, with world-volume couplings to the branes
given by
\beqa
i \mu_9 \frac{1}{2!}\int_{D9_a} C_6 \wedge \cf_a^2, \nonumber \\ 
i \mu_9 \frac{1}{4!}\int_{D9_a} C_2 \wedge \cf_a^4, \nonumber 
\label{10Dfields}
\eeqa
where $\cf_a = 2\pi \a^\prime F_a + B$ is the gauge invariant 
two-form flux living on brane $a$, and 
$\mu_9 = (2\pi)^{-9} M_s^{10}$ is the charge of the 
D9-brane under RR fields. When going to four dimensions,
(\ref{FFcouplings}) couplings will arise by dimensional reduction.
In \cite{imr} these couplings were computed for a factorizable 
brane $a$, and they turn out to be
\beqa
n_a^1 n_a^2 n_a^3 \frac{i}{4\pi}\int_{M_4} C^0 \wedge  F_a \wedge F_a, 
\nonumber \\ 
n_a^I m_a^J m_a^K \frac{i}{4\pi}\int_{M_4} C^I \wedge  F_a \wedge F_a, 
\label{Cfields}
\eeqa
where the $n'$s and $m'$s stand for (fractional) wrapping numbers on each 
torus. $C^0, C^I$ are four-dimensional scalar RR fields defined as
\beqa
C^0 = (4\pi^2 \a^\prime)^{-3} \int_{T^6} C_6, \ \
C^I = (4\pi^2 \a^\prime)^{-1}\int_{T^2_I} C_2.
\label{4Dfields}
\eeqa
 
From these couplings we see 
that the imaginary part of the gauge kinetic function should be of the form
\beq
{\rm Im}(f_a) = \frac{1}{4\pi} \left( n_a^1 n_a^2 n_a^3 C^0 
+ \sum_I n_a^I m_a^J m_a^K  C^I \right),
\label{imaginary}
\eeq
whereas the real part should depend only on the volume of the brane
\beq
{\rm Re}(f_a) = {M_s^3 \over 2\pi \lambda} 
\sqrt{\prod_{i=1}^3 \left(\left(n_a^iR_1^{(i)}\right)^2+
\left(m_a^iR_2^{(i)}\right)^2\right)}.
\label{real}
\eeq

Expressions (\ref{imaginary}) and (\ref{real}) should be the real and 
imaginary part of the same holomorphic function, though they seem very 
different. (\ref{imaginary}) has a linear dependence on four RR fields, 
whereas (\ref{real}) depends non-linearly on the NSNS moduli describing 
the complex structure of the torus. The solution to this apparent puzzle 
comes from the fact that the D6-brane $a$ not always forms a 
supersymmetric system by itself,
but has to preserve a supersymmetry with the O6-plane, lying on the homology 
cycle $[(1/\beta^1,0)(1/\beta^2,0)(1/\beta^3,0)]$. If this brane and the 
O6-plane (or equivalently, if this brane and its mirror image $a^*$) 
preserve a supersymmetry, then the length of this brane can be expressed 
by a sum of fields, just as in (\ref{imaginary}). If, for instance, 
the twist vector with respect to the orientifold is given by 
$v_\vt = (\vt^1, \vt^2, \vt^1+\vt^2, 0)$, $\vt^i > 0$, 
then by some basic trigonometry it can be shown that the volume 
of this brane $a$ can be expressed as:
\beqa
{\A l_a \A \over (2\pi)^3} & = n_a^1  n_a^2 n_a^3 R_1^{(1)} R_1^{(2)} R_1^{(3)}
+ n_a^1 m_a^2 m_a^3 R_1^{(1)} R_2^{(2)} R_2^{(3)} \nonumber \\
& + m_a^1 n_a^2 m_a^3 R_2^{(1)} R_1^{(2)} R_2^{(3)} 
- m_a^1 m_a^2 n_a^3 R_2^{(1)} R_2^{(2)} R_1^{(3)}
\label{length1}
\eeqa

In order to properly compare both real and imaginary parts of the 
gauge kinetic function, let us first translate expression 
(\ref{length1}) into its T-dual counterpart. For this we must 
apply the T-duality transformations 
\beqa
R_2^{(I)} \leftrightarrow {\a^\prime \over R_2^{(I)}} \\
R_1^{(I)} \leftrightarrow R_1^{(I)} \\
\lambda \leftrightarrow 
{\a^{\prime \ 3/2} \lambda \over \prod_I R_2^{(I)}}.
\label{Tdual}
\eeqa
After such transformations we can express the real part of the gauge 
kinetic function entirely in terms of a sum of NSNS fields
\beq
{\rm Re}(f_a)  = n_a^1  n_a^2 n_a^3 S_f 
+ n_a^1 m_a^2 m_a^3  U^1_f 
 + m_a^1 n_a^2 m_a^3 U^2_f 
- m_a^1 m_a^2 n_a^3 U^3_f,
\label{real2}
\eeq
where we have defined
\beqa
S_f & \equiv  & {M_s^6 \over 2\pi\lambda} \prod_{I=1}^3 R_1^{(I)} R_2^{(I)},
\label{NSflux1} \\
U^I_f & \equiv  & {M_s^2 \over 2\pi\lambda}  R_1^{(I)} R_2^{(I)}.
\label{NSflux2}
\eeqa
Note that these correspond to the standard 4-D dilaton of toroidal 
compactifications and the three Kahler moduli of the three tori
\cite{ssr}.
The subscript $f$ stands for the fluxes 
(D9-brane) picture, where these NSNS
four-dimensional moduli are relevant. In the T-dual picture of branes 
at angles, these same four-dimensional fields take the form
\beqa
S_a & \equiv  & {M_s^3 \over 2\pi\lambda} \prod_{I=1}^3 R_1^{(I)},
\label{NSangles1} \\
U^I_a & \equiv  & {M_s^3 \over 2\pi\lambda}  R_1^{(I)} R_2^{(J)} R_2^{(K)}.
\label{NSangles2}
\eeqa
Notice that from a four-dimensional viewpoint, we cannot distinguish
from which of the T-dual systems we are compactifying. Fields 
(\ref{NSflux1}, \ref{NSflux2}) and 
(\ref{NSangles1}, \ref{NSangles2}) are 
exactly the same when talking about low energy physics 
in four dimensions. We can, then, delete the subscript 
$a$ or $f$ in order to describe our field theory. We should only 
take into account which field are we working with when doing
a computation where the geometry of the compactification
is relevant. Any of the T-dual choices should give us the 
same result.

It is now easy to express the gauge kinetic function as an holomorphic 
function on the relevant fields, namely as a sum of four complex fields, 
whose real part consist of some complex structure field and the imaginary 
part of some RR untwisted field. In general, we find that given a preserved 
$\cn= 1$ SUSY described by the vector 
$\tilde r = \oh (\eps^1, \eps^2, \eps^3, \eps^4)$ belonging to
(\ref{r}), then the gauge kinetic function can be expressed as:
\beq
f_a = n_a^1 n_a^2 n_a^3 \tilde S 
+ \sum_I n_a^I m_a^J m_a^K  \tilde U^I,
\label{gkf}
\eeq
where the complex fields $\tilde S, \tilde U$ are given by
\beqa
\tilde S & = & S + {i \over 4\pi} C^0, 
\label{matching1} \\
\tilde U^I & = & - U^I \eps^J \eps^K + {i \over 4\pi} C^I.
\label{matching2}
\eeqa

Thus, we have defined four complex scalar fields that will 
appear in our effective field theory description of our 
compactification. It is interesting to notice that the definition 
of these fields depends on which specific supersymmetry is 
preserved by our brane $a$ and its mirror. Indeed, departing
from the $\cn = 4$ bulk supersymmetry preserved by our plain
Type I theory, we have defined four $\cn = 1$ independent 
subalgebras represented by four independent spinors (\ref{r}).
Each of these $\cn = 1$ SUSY's can be expressed in terms of 
one of these vectors $\tilde r$, in turn encoded in terms of
the $\eps$'s appearing in (\ref{matching2}). 
To sum up, given a brane
preserving a $\cn = 1$ SUSY with its mirror, we can express
the gauge kinetic function and the Kahler potential (which 
define the supersymmetric low energy effective action) in 
terms of some complex scalar fields. The form of these 
fields is inherited by bulk superfields,
as expected, but with some relative signs 
$\eps^i$ depending on the specific $\cn = 1$ preserved by 
the brane. We will call each  choice of signs in 
(\ref{matching2}) a different SUSY prescription, that
tell us how RR and NSNS bulk field relate in order to enter
a chiral $\cn = 1$ multiplet as a complex scalar field.
Notice that each of the branes appearing in our hexagonal 
models from Section 4 do not only preserve $\cn = 1$, but 
$\cn = 2$. Then, two different $\tilde r$ vectors could be 
considered. Our complex superfields are so described by two 
different SUSY prescriptions which are, in fact, the same,
since only one of the fields $\tilde U^I$ couples to the 
brane.

\subsection{Fayet-Iliopoulos terms}

We saw in the previous section how for particular choices of the
complex structure moduli one can obtain an 
unbroken $\cn =1$ SUSY at an intersection. Now, 
if we modify slightly the value of those moduli supersymmetry
will be broken. It is reasonable to expect that the effect
of this breaking will be approximately described by the 
turning on of a Fayet-Iliopoulos term in the theory. We will 
describe now in some detail how this happens and  how the
scalar fields at the intersection get masses from the FI-terms.
Those masses agree as expected with the ones obtained 
from  the string mass formulae eq.(\ref{scalars}).

In the previous subsection we 
also showed  the gauge kinetic function that
should
describe the effective field theory of a D6-brane preserving a supersymmetry 
with the O6-plane. If we define a twist vector $v_{\vt}$ between the 
orientifold and this brane, the condition for preserving such supersymmetry 
can be reformulated as that one of the `scalars' in (\ref{scalars}) should 
become massless\footnote{Such scalars are in fact non-existent in the low 
energy theory, since no open string lives on the intersection of a brane and 
the orientifold plane. We could, in turn, consider the scalars arising in the 
$D6_aD6_{a^*}$ sector, whose twist vector is $2v_{\vt}$ and thus have two 
times the mass of these fictitious scalars. Strictly speaking, all the 
considerations
made in this section regarding the system brane-orientifold should be 
translated to the system brane-mirror brane.}, thus defining a SUSY wall
in the 3-dimensional parameter space $(\vt^1,\vt^2,\vt^3)$. Since we have 
four different scalars, four SUSY walls appear, forming a tetrahedron.
The structure of this tetrahedron and its physical consequences have 
been studied in \cite{imr,torons}. Briefly stated, when standing inside
the tetrahedron all of the scalars (\ref{scalars}) are massive, signaling
that no supersymmetry is preserved by both branes. When standing outside 
this tetrahedron one of such scalars will become tachyonic, and this 
indicates that the system is unstable as it stands.

When a system composed of two factorizable branes $a$ and $b$ lies in 
one of these tetrahedron walls they are  in a marginal 
stability (MS) wall. The concept of marginal stability is in fact more 
general, and can be applied to branes wrapping SLAGs in general cycles 
\cite{Denef}. As  discussed in  \cite{kg}, when 
crossing such a wall a tachyon may appear at the intersection of the 
two branes, signaling an unstability against the recombination of these 
two branes into a single one. This final brane will wrap a SLAG whose 
homology class is determined by the sum of the RR charges of the two 
previous branes. Supersymmetry will be restored under recombination, 
while one of the $U(1)$'s will become massive. At the other side of the 
MS wall, this recombination is not favoured locally in terms of difference 
of tensions, so the pair of branes will not recombine into one. We will 
then have a non-supersymmetric configuration with gauge group 
$U(1)_a \t U(1)_b$. As noted in \cite{kg}, this behaviour can be understood 
from a Field Theory viewpoint by including a Fayet-Iliopolous term 
$D \xi$ in the effective lagrangian, so that the potential energy becomes
\beq
V_{FI}(\phi) = {1 \over 2 g^2} (|\phi|^2 + \xi)^2,
\label{potentialFI}
\eeq
where $\xi$ is the transversal separation from the MS wall, and $\phi$ 
are the lightest complex scalars living at the intersection. 
For a T-dual description of this phenomenon see \cite{Witten}.

When considering a full configuration of branes, where several FI terms 
arise, we would expect several contributions to the potential energy,
each of them given by
\beq
V_{FI, a}(\phi_i) = {1 \over 2 g_a^2} (\sum_i q_a^i|\phi_i|^2 + \xi_a)^2.
\label{potentialFI2}
\eeq
Here, $\xi_a$ represents the FI-term associated to $U(1)_a$, and
$\phi_i$ are all the scalar fields charged under $U(1)_a$ with charge
$q_a^i$. It should be possible to compute these FI terms from an 
effective field theory description of the gauge theory living on these 
branes. Let us consider, as before, an orientifold 
compactification and a brane $a$ preserving a SUSY $\tilde r$ with the 
O6-plane. When separating a bit from the SUSY wall, the supersymmetric field 
theory should break. This supersymmetry breaking should be understood as some 
FI-terms in this theory. As noted in \cite{Dine}, the FI terms in a 
SUSY theory can be deduced from the couplings of the form
\beq
{Q_a^i \over 2\pi \a^\prime} \int_{M_4} B_2^i \wedge F_a,
\label{Fcouplings}
\eeq
that arise in the compactified four-dimensional theory.
Here $B_2^i$ are   four-dimensional antisymmetric fields
(dual to the RR C-scalars discussed above) 
and the dimensionless coeficcient $Q_a^i$ represents 
how the brane $a$ couples
to such field. These couplings are the mediators of  
the Green-Schwarz anomaly cancellation mechanism 
\footnote{Note that  massless anomaly-free $U(1)$'s will have 
no FI-terms. However anomaly-free $U(1)$'s which 
become massive through a $B\wedge F$ coupling may in
general get a FI away from the SUSY wall.}. 
Just as done above, we can compute which couplings of type 
(\ref{Fcouplings}) appear from a four-dimensional viewpoint
when dimensionally reducing the couplings (\ref{10Dfields}).
As found  in \cite{imr}, they turn out to be
\beqa
{1 \over 4\pi^2 \a^\prime} N_a m_a^1 m_a^2 m_a^3
\int_{M_4} B_2^0 \wedge  F_a, 
\nonumber \\ 
{1 \over 4\pi^2 \a^\prime} N_a m_a^I n_a^J n_a^K 
\int_{M_4} B_2^I \wedge  F_a, 
\nonumber
\eeqa
where the $N_a$ factors arise from the  normalization 
of the $U(1)_a$ subgroup of $U(N_a)$ (see \cite{iru,afiru}).

These two-forms $B_2$ living in our four-dimensional field theory 
are again defined by partially integrating the couplings 
(\ref{10Dfields}) on $T^6$
\beqa
B_2^0 = C_2, \ \
B_2^I = (4\pi^2\a^\prime)^{-2} \int_{(T^2_J)\t(T^2_K)} C_6.
\label{4Dfields2}
\eeqa
Both scalars (\ref{4Dfields}) and two-forms (\ref{4Dfields2}) 
are four-dimensional RR fields related to each other by hodge 
duality in four dimensions:
\beqa
dC^0 = - * dB_2^0 \\
dC^I = - * dB_2^I.
\label{hodge}
\eeqa

By supersymmetric field theory arguments, we would expect 
Fayet-Iliopolous terms of the form 
\beq
 D_a {\xi_a \over g_a^2} = D_a 
\left({\p\ck \over \p V_a} \right)_{V = 0},
\label{Fayet}
\eeq
where $V_a$ is the vector multiplet associated to the 
massive $U(1)_a$, 
and $\ck$ the Kahler potential, whose gauge invariant expression
 in these toroidal compactifications is given by
\beq
\ck =  {M_{Planck}^2 \over 8\pi} \left(- log \left(\tilde S + \tilde S^*
- \sum_a Q_a^0 V_a \right) - 
\sum_{i=1}^3 log \left(\tilde U^i + \tilde U^{i *} - \sum_a Q_a^i V_a
\right) \right).
\label{Kpotential}
\eeq

Substituting in (\ref{Fayet}) under the SUSY prescription 
(\ref{matching1}, \ref{matching2}) 
we obtain 
\beqa
{\xi_a \over g_a^2}  & = & {M_{P}^2  \over 32\pi^2} 
{(2\pi)^3 \lambda N_a \over {\rm Vol} (T^6) M_s^3}\prod_{i=1}^3 \eps^i 
 \left((2\pi)^3 \prod_{i=1}^3 \eps^i m_a^i R_2^{(i)} 
(2\pi)^3 \sum_{i=1}^3 \eps^i m_a^i n_a^j n_a^k 
R_2^{(i)} R_1^{(j)} R_1^{(k)} \right) \nonumber \\
 & = & {M_s^{5} N_a \over (2\pi)^5 \lambda} \prod_{i=1}^3 \eps^i 
 \left((2\pi)^3 \prod_{i=1}^3 \eps^i m_a^i R_2^{(i)} 
- (2\pi)^3 \sum_{i=1}^3 \eps^i m_a^i n_a^j n_a^k 
R_2^{(i)} R_1^{(j)} R_1^{(k)} \right), 
\label{Fayet2}
\eeqa
where we have used $M_P^2 = 8 M_s^{8}{\rm Vol}(T^6)/ (2\pi)^6\lambda^2$
\cite{ssr}.
For convenience we  have written  this expression in terms of
``D6-branes at angles'' geometric moduli,
One can check that 
the term in brackets vanishes at the SUSY wall, thus giving the 
expected behaviour for a FI-term. It turns out that it has a 
simple dependence on the separation parameter $\d$ from the 
supersymmetric case. To illustrate this, let us take our 
previous example and variate the complex structures in order 
to have a twist vector $v_\vt = (\vt^1, \vt^2, \vt^1+\vt^2 + \d, 0)$, 
$\vt^i > 0$. Then, for small $\d$, our approximate SUSY is still given by
the vector $\tilde r = \oh (+,+,-,-)$ and after some trigonometry we find 
that our expression becomes
\beq
{\xi_a \over g_a^2} = - {M_s^{5} \over (2\pi)^5 \lambda} 
N_ a\A l_a \A \ sin (\pi\d) = 
- {sin (\pi\d) \over (2\pi) g_a^2} M_s^{2}
\label{Fayet3}
\eeq
where we have used (\ref{coupling}), and taken into account that
$g^2_a \equiv g^2_{U(1)_a}$ (see footnote 11). In the 
limit of small separation from the SUSY wall, that is when 
$\d \ll 1 $, is where we expect our field theory approximation 
to be valid. In this limit we can approximate our FI-term by
\beq
\a^\prime \xi_a \approx - {\d \over 2}.
\label{Fayet4}
\eeq

In order to compute the mass of a scalar living at the intersection of two
branes, let us take two D6-branes $D6_a$ and $D6_b$ whose separation from 
the same SUSY wall is given by $\d_a$ and $\d_b$, respectively. 
By looking at the effective potential (\ref{potentialFI2}), we would 
expect the mass of this scalar to be given by
\beq
\a^\prime m_{ab}^2 = \a^\prime(-q_a \xi_a - q_b \xi_b) \approx 
{1 \over 2} (\d_a - \d_b) \ .
\label{FImass}
\eeq
Notice that this result is in agreement with the masses for the
scalars lying at a intersection obtained from the string mass 
formulae in  (\ref{scalars}).
Thus we see that, for small deviations from a SUSY configuration,
the masses of the scalars at an intersection may be understood as 
coming from a Fayet-Iliopoulos term.

\subsection{Application to  $\cn = 1$ SUSY models}

Let us now apply our Field Theory results to a $\cn = 1$ SUSY model, 
as those presented in \cite{csu}. Recall that in those 
$\IZ_2 \t \IZ_2$ models branes were related to the O6-plane by a 
rotation from the same subgroup $SU(3) \subset SU(4)$. 
Namely, they had the twist vectors (\ref{vectors}) which, in 
our hexagonal construction from figure \ref{hexagon}, 
translates into a restriction of such general models to 
those containing the SUSY triangle formed by $a_2$, $b_2$ 
and $c_2$ branes, who share the SUSY $\tilde r_4 = \oh (+,+,+,+)$.
Using formulae (\ref{matching1}, \ref{matching2}), we see that 
the volume of any of such branes can be expressed as
\beq
\Arrowvert l_a \Arrowvert = \prod_{i=1}^3 n_a^i R_1^{(i)} 
- \sum_{i=1}^3 n_a^i m_a^j m_a^k R_1^{(i)} R_2^{(j)} R_2^{(k)}.
\label{length2}
\eeq

Let us consider then a full configuration of D6-branes preserving this 
 $\cn = 1$ SUSY with the orientifold plane. We can again look at the 
potential derived from the DBI action by dimensional reduction
\beq
V = {T_6 \over \lambda} 
\left( \sum_a N_a \A l_a \A -  \A l_{ori} \A\right),
\label{potential}
\eeq
where the contribution from mirror branes need not be taken into 
in this particular computation. Since we are dealing with a 
supersymmetric configuration this quantity should vanish in a 
consistent compactification. In these particular $\IZ_2 \t \IZ_2$ 
models, the negative tension of the orientifold planes is given by
\beq
\Arrowvert l_{ori} \Arrowvert = 16 \prod_{i=1}^3  R_1^{(i)} 
+ 16 \sum_{i=1}^3 R_1^{(i)} \beta^j R_2^{(j)} \beta^k R_2^{(k)}.
\label{lengthori}
\eeq
The conditions for having no NS tadpoles simply translate into a vanishing 
potential, so they are
\beqa
& \sum_a N_a \ n_a^1 n_a^2 n_a^3 = 16 \\
& \sum_a N_a \ n_a^i m_a^j m_a^k = -16 \beta^j \beta^k
\label{tadpoles4}
\eeqa
which are just the RR tadpoles cancellation conditions 
derived in \cite{csu}.
The results from previous sections also apply to this kind of 
models, having a gauge kinetic function for each brane 
whose holomorphicity or SUSY prescription 
(\ref{matching1}, \ref{matching2}) is the same for every brane. 

As a second application, let us rederive the FI terms appearing in
(\ref{potentialFI2}) from a different viewpoint. The contribution 
to this effective potential coming from
$\xi_a^2 / g_a^2$ should come from the closed strings modes that 
participate 
in (\ref{potential}). Indeed, we should be able to compute this 
contribution by considering the value of (\ref{potential})
slightly away from the SUSY wall. Now, notice that a $\cn = 1$ SUSY 
model where the complex structures have been varied still satisfies
\beq
\Arrowvert l_{ori} \Arrowvert
= \sum_a N_a \Arrowvert l_a \Arrowvert \ {\rm cos}\ \pi\d_a
\label{lengthori2}
\eeq
where $\d_a$ is the separation 
of each brane from the SUSY wall. The potential (\ref{potential}) is then 
given by
\beq
V = {T_6 \over \lambda} \sum_a N_a \A l_a \A
\left( 1 - \ {\rm cos}\ \pi\d_a \right) = 
\frac{M_s^4}{(2\pi)^2}\sum_a {1 \over g_{U(1)_a}^2}
\left( 1 - \ {\rm cos}\ \pi\d_a \right),
\label{potential2}
\eeq
where we have used the explicit form of $T_6$ and formula (\ref{coupling}).
For small separation we again can approximate this expression, obtaining
\beq
V \approx
\oh \sum_a {1 \over g_{U(1)_a}^2} \left({\d_a\over 2 \a^\prime}\right)^2,
\label{potential3}
\eeq
which again gives the expected behaviour for a 
potential from a FI term.

Finally, let us study how the system behaves when the $\cn = 1$ 
supersymmetry of these kind of models is slightly broken. 
We will focus on a fairly general subsector of a SUSY (or Q-SUSY) 
compactification where we have  a SUSY triangle
with branes of type $a_2,b_2$ and $c_2$ only. 
In order to slightly break $\cn = 1$ SUSY, we can make a small 
variation on the complex structure. Let us, for instance, 
place ourselves in a point of 
the moduli space of complex structures where the SUSY condition 
is satisfied. We can make a small variation on the radii of the 
first  and second tori, which induces a small variation on the
angles the branes make. Such variations are parametrized by
small $\d$'s defined as $\vt_a^2 = \vt_a + \d^2$, and 
$\vt_b^1 = \vt_b + \d^1$. Notice that this already fixes the 
$c$ sector departure from SUSY breaking. In general, we 
can parametrize the small Q-SUSY breaking by these two terms 
$\d^i$, $i = 1,2$. Table \ref{susys} is now modified to
\TABLE{\renewcommand{\arraystretch}{1.2}
\begin{tabular}{|c|c|c|}
\hline
 Brane & Twist vector & approx. SUSY \\
\hline
\hline 
$a_2$, ($a_2^*$) & $\pm(0,\vt_a+\d^2,-\vt_a)$ & $\tilde r_1$, $\tilde r_4$ \\
\hline 
$b_2$, ($b_2^*$) & $\pm(-\vt_b-\d^1,0,\vt_b)$ & $\tilde r_2$, $\tilde r_4$ \\
\hline 
$c_2$, ($c_2^*$) & $\pm(\vt_c+\a^1\d^1,-\vt_c-\a^2\d^2,0)$ & 
$\tilde r_3$, $\tilde r_4$  \\
\hline 
\end{tabular}
\label{susybreaking1}
\caption{\small Small SUSY breaking for a $\cn = 1$ triangle.}}

In table \ref{susybreaking1}, $\a^1$ and $\a^2$ are two proportionality 
factors arising from the different length of the branes. 
Their expression, valid for small $\d$, is
given by $\a^1 = \frac{{\rm sen}2\vt_c}{{\rm sen}2\vt_b}$ and
$\a^2 = \frac{{\rm sen}2\vt_c}{{\rm sen}2\vt_a}$. Let us fix a hierarchy
on the angle's values. Take, for instance, $\vt_a < \vt_b < \vt_c$. 
We can easily compute the mass spectrum arising 
from the lightest scalars living at the intersections. This is done in 
table \ref{susybreaking2}, where we show the corresponding twist vector, 
and we compute the square mass of such scalar particles, 
the superpartners of each massless fermion living at these 
intersections.
Recall that, as shown above, these masses may be understood as coming
from induced FI-terms for the three $U(1)$ fields in these models. 
\TABLE{\renewcommand{\arraystretch}{1.2}
\begin{tabular}{|c|c|c|c|}
\hline
\hline
Intersection & Twist vector & approx. SUSY & mass$^2$ \\
\hline
\hline
$a_2b_2$ & $(-\vt_b-\d^1,-\vt_a-\d^2,\vt_b+\vt_a)$ &
$\tilde r_4$ & $\oh(\d^1+\d^2)$ \\
$a_2b_2^*$ & $(\vt_b+\d^1,-\vt_a-\d^2,-\vt_b+\vt_a)$ &
$\tilde r_4$ & $\oh(-\d^1+\d^2)$ \\
\hline
$a_2c_2$ & $(\vt_c+\a^1\d^1,-\vt_c-\vt_a-(\a^2+1)\d^2,\vt_a)$ &
$\tilde r_4$ & $\oh(\a^1\d^1-(\a^2+1)\d^2)$ \\
$a_2c_2^*$ & $(-\vt_c-\a^1\d^1,\vt_c-\vt_a+(\a^2-1)\d^2,\vt_a)$ &
$\tilde r_4$ & $\oh(-\a^1\d^1+(\a^2-1)\d^2)$ \\
\hline
$b_2c_2$ & $(\vt_c+\vt_b+(\a^1+1)\d^1,-\vt_c-\a^2\d^2,-\vt_b)$ &
$\tilde r_4$ & $\oh(-(\a^1+1)\d^1+\a^2\d^2)$ \\
$b_2c_2^*$ & $(-\vt_c+\vt_b-(\a^1-1)\d^1,\vt_c+\a^2\d^2,-\vt_b)$ &
$\tilde r_4$ & $\oh((\a^1-1)\d^1-\a^2\d^2)$ \\
\hline
\hline
\end{tabular}
\label{susybreaking2}
\caption{\small Small SUSY breaking at intersections. 
$\cn = 1$ triangle.}}

Notice that, for any value of the SUSY breaking
parameters $\d^1$, $\d^2$ a tachyon always appears at some intersection.
This can be easily seen if we notice that, for this hierarchy of angles
($\vt_a^j < \vt_b^j < \vt_c^j$), identities between sparticles 
masses such as the following hold:
\beq
m^2(a_2b_2) + m^2(a_2b_2^*) + m^2(a_2c_2) + m^2(a_2c_2^*) = 0,
\label{sparticles}
\eeq
which clearly implies that one of the scalars living at these four 
intersections must become tachyonic when breaking the $\cn = 1$
SUSY by varying the complex structure. It is important to stress
that a general $\cn = 1$ model does not have necessarily to 
include such a triangle. However, the appearance of a tachyon
at some intersection when slightly varying the complex structure
from the supersymmetric case seems a general feature of any
$\cn = 1$ compactification satisfying tadpole cancellation.

Note that these tachyons are nothing but an indication that 
the initial configuration of branes is unstable and that the
two intersecting branes will fuse into a single one minimizing the
energy
\cite{csu}. In the process the gauge symmetry will 
be broken and the rank reduced. 
Thus, as usually happens in all  known $D=4$
string constructions the presence of a FI-term does not signal 
SUSY-breaking but just the existence of a nearby vacuum 
which is again supersymmetric.
We will now see that in specific Q-SUSY models that is not in general the
case, i.e., the presence of FI-terms does not necessarily give rise to
tachyonic masses for any scalars and hence SUSY is actually broken 
at the intersections.

\subsection{Q-SUSY models: NS-tadpoles and FI-terms}

Unlike the $\cn = 1$ case, the scalar potential for the $S$,$U^I$ 
Neveu-Schwarz fields does not vanish in Q-SUSY models, i.e., 
there are uncancelled NS tadpoles. 
This is expected, the theory
being non-supersymmetric. However, one can see that the 
form of this scalar potential is particularly simplified in the case of 
Q-SUSY models, compared to a general non-SUSY toroidal model.
In particular, we find that the potential is linear in 
the $S,U$ NS fields and for some simple cases the four-dimensional dilaton 
($S$) tadpole even vanishes once the RR-tadpoles vanish. 
Let us first check this point in a simple example.

Let us consider a subclass of Q-SUSY
models where no branes of type $c$ appear. We will call such
models quadrilateral or square quiver models 
(see fig.\ref{cuadrado} and next section). To be concrete, we
will consider a configuration where four different branes appear
each of a different kind, as shown in table \ref{qmodel2}.
\TABLE{\renewcommand{\arraystretch}{1.5}
\begin{tabular}{|c||c|c|c||c|}
\hline 
 $N_i$ & $(n_i^1,m_i^1)$ & $(n_i^2,m_i^2)$ & $(n_i^3,m_i^3)$ & $v_i$ \\
\hline
\hline $N_{a_1}$ & $(1/\beta^1,0)$  &  $(n_{a_1}^2,m_{a_1}^2)$ &
 $(n_{a_1}^3,m_{a_1}^3)$ & $(0,\vt_{a_1}^2,\vt_{a_1}^3)$ \\
\hline $N_{a_2}$ & $(1/\beta^1,0)$  & $(n_{a_2}^2,m_{a_2}^2)$ &
 $(n_{a_2}^3,-m_{a_2}^3)$ & $(0,\vt_{a_2}^2,-\vt_{a_2}^3)$ \\
\hline $N_{b_1}$ & $(n_{b_1}^1,m_{b_1}^1)$ & $ (1/\beta^2,0)$  &
$(n_{b_1}^3,m_{b_1}^3)$ & $(\vt_{b_1}^1,0,\vt_{b_1}^3)$ \\
\hline $N_{b_2}$ & $(n_{b_2}^1,-m_{b_2}^1)$ & $ (1/\beta^2,0)$  &
$(n_{b_2}^3,m_{b_2}^3)$ & $(\vt_{b_2}^1,0,\vt_{b_2}^3)$ \\
\hline \end{tabular}
\label{qmodel2}
\caption{\small Example of D6-brane wrapping numbers giving rise to
a square quiver Q-SUSY model. We are again taking 
$n_i^j, m_i^j \geq 0$, $i = a, b$, $j = 1, 2, 3$.}}

We will suppose that this is a RR tadpole-free configuration,
which amounts to the following restrictions
\beq
\begin{array}{c}
\sum_{i=1}^2 \left( \frac{1}{\beta^1}N_{a_i}n_{a_i}^2n_{a_i}^3  
+ \frac{1}{\beta^2}N_{b_i}n_{b_i}^1n_{b_i}^3 \right) = 16
\\
N_{a_1}m_{a_1}^2m_{a_1}^3 = N_{a_2}m_{a_2}^2m_{a_2}^3  \\
N_{b_1}m_{b_1}^1m_{b_1}^3 = N_{b_2}m_{b_1}^2m_{b_2}^3
\end{array}
\label{tadpoles5}
\eeq

The length of the branes composing such a model is easily computed 
from (\ref{matching1}, \ref{matching2}) 
\beqa
& & \A l_{a} \A = (2\pi)^3 \frac{1}{\beta^1}\left(n_{a}^2 n_{a}^3 
R_1^{(1)}R_1^{(2)}R_1^{(3)} +  
m_{a}^2 m_{a}^3 R_1^{(1)}R_2^{(2)}R_2^{(3)} \right) 
\nonumber \\
& & \A l_{b} \A = (2\pi)^3 \frac{1}{\beta^2}\left(n_{b}^1 n_b^3
R_1^{(1)}R_1^{(2)}R_1^{(3)} + 
m_b^1 m_b^3 R_2^{(1)}R_1^{(2)}R_2^{(3)} \right),
\nonumber 
\label{qlength}
\eeqa
where we index $a = a_1, a_2$, same for $b$. When substituting these 
quantities in the potential (\ref{potential}), we get 
(again not including  mirror branes)
\beqa
 V &  = & {T_6 \over \lambda} 
\left( \sum_a N_a \A l_a \A - \A l_{ori} \A\right) \nonumber \\
& = & {M_s^7 \over (2\pi)^3 \lambda}  
\left( 
\frac{1}{\beta^1}N_{a}{m_{a}^2 m_{a}^3} R_1^{(1)}R_2^{(2)}R_2^{(3)} +  
\frac{1}{\beta^2}N_{b}{m_{b}^1 m_{b}^3} R_2^{(1)}R_1^{(2)}R_2^{(3)}
\right) \nonumber \\
& = & {M_s^4 \over (2\pi)^2} 
\left(
\frac{1}{\beta^1}N_{a}{m_{a}^2m_{a}^3} U^1 +
\frac{1}{\beta^2}N_{b}{m_{b}^1m_{b}^3} U^2 
\right).
\label{NSpotential3}
\eeqa
Notice that, as promised, the dependence of $U^1$, $U^2$ is linear on
these fields. This is a general characteristic of Q-SUSY models
and not of this particular example. Note also that only two out of
four NSNS fields appear in brackets, $S$ and $U^3$ being absent.
There is no NS tadpole for the $D=4$ dilaton field $S$, and there is
only one NS tadpole left corresponding to a linear 
combination of the $U^1$ and $U^2$ fields.  Thus, the structure of NS
tadpoles in this class of models is substantially simplified
compared to generic non-SUSY orientifold models.

Let us now do the same exercise we did for the case of
$\cn =1$ models concerning FI-terms but for this Q-SUSY
configuration.
We will  show how in this Q-SUSY example turning on a FI-term 
does not necessarily lead to gauge symmetry breaking (but unbroken
SUSY) as happenned in the $\cn=1$ case. Rather one can get unbroken
gauge symmetry (no tachyons) but broken SUSY at the intersections.

In order to study how the system behaves when the Quasi-supersymmetry is 
slightly broken, we can make a small variation of the complex structures
just as we did with the SUSY triangle. Thus, we will first consider
a complex structure such that $\vt_{a_i}^2 = \vt_{a_i}^3$  and
$\vt_{b_i}^1 = \vt_{b_i}^3$, $i = 1,2$ and then perform a small change on 
the quotient of radii such that there is small departure from 
these equalities, again parametrized by $\d$'s. The corresponding 
twist vectors are shown in table \ref{qsusybreaking1}.
\TABLE{\renewcommand{\arraystretch}{1.2}
\begin{tabular}{|c|c|c|}
\hline
 Brane & Twist vector & approx. SUSY \\
\hline
\hline 
$a_1$, ($a_1^*$) & $\pm(0,\vt_{a_1}+\d^2,\vt_{a_1})$ & 
$\tilde r_2$, $\tilde r_3$ \\
\hline 
$a_2$, ($a_2^*$) & $\pm(0,\vt_{a_2}+\a^2\d^2,-\vt_{a_2})$ & 
$\tilde r_1$, $\tilde r_4$ \\
\hline 
$b_1$, ($b_1^*$) & $\pm(\vt_{b_1}+\d^1,0,\vt_{b_1})$ & 
$\tilde r_1$, $\tilde r_3$ \\
\hline 
$b_2$, ($b_2^*$) & $\pm(-\vt_{b_2}-\a^1\d^1,0,\vt_{b_2})$ & 
$\tilde r_2$, $\tilde r_4$ \\
\hline 
\end{tabular}
\label{qsusybreaking1}
\caption{\small Small Q-SUSY breaking for a square quiver model.
In this general example we are allowing $\vt_{b_1} \neq \vt_{b_2}$. 
There are some proportionality factors defined as
$\a^1 \equiv  \frac{{\rm sen}2\vt_{b_2}}{{\rm sen}2\vt_{b_1}}$,
$\a^2 \equiv  \frac{{\rm sen}2\vt_{a_2}}{{\rm sen}2\vt_{a_1}}$.}}

Just as in the SUSY triangle, we can again compute the masses of the 
lightest scalars at the eight relevant intersections. For this 
will consider again   a hierarchy of angles, which we choose to be 
$\vt_{a_1} < \vt_{a_2} < \vt_{b_1} < \vt_{b_2}$. The square mass 
for the corresponding qsuperpartners of the massless fermions
is computed in table \ref{qsusybreaking2}.

\TABLE{\renewcommand{\arraystretch}{1.2}
\begin{tabular}{|c|c|c|c|}
\hline
\hline
Intersection & Twist vector & approx. SUSY & mass$^2$ \\
\hline
\hline 
$a_1b_1$ & $(\vt_{b_1}+\d^1,-\vt_{a_1}-\d^2,\vt_{b_1}-\vt_{a_1})$ & 
$\tilde r_3$ & $\oh(\d^1-\d^2)$ \\
$a_1b_1^*$ & $(-\vt_{b_1}-\d^1,-\vt_{a_1}-\d^2,-\vt_{b_1}-\vt_{a_1})$ & 
$\tilde r_3$ & $\oh(\d^1+\d^2)$\\
\hline 
$a_1b_2$ & $(-\vt_{b_2}-\a^1\d^1,-\vt_{a_1}-\d^2,\vt_{b_2}-\vt_{a_1})$ &
$\tilde r_2$ & $\oh(-\a^1\d^1+\d^2)$\\
$a_1b_2^*$ & $(\vt_{b_2}+\a^1\d^1,-\vt_{a_1}-\d^2,-\vt_{b_2}-\vt_{a_1})$ & 
$\tilde r_2$ & $\oh(\a^1\d^1+\d^2)$\\
\hline
$a_2b_1$ & $(\vt_{b_1}+\d^1,-\vt_{a_2}-\a^2\d^2,\vt_{b_1}+\vt_{a_2})$ &
$\tilde r_1$ & $\oh(\d^1+\a^2\d^2)$\\
$a_2b_1^*$ & $(-\vt_{b_1}-\d^1,-\vt_{a_2}-\a^2\d^2,-\vt_{b_1}+\vt_{a_2})$ &
$\tilde r_1$ & $\oh(-\d^1+\a^2\d^2)$\\
\hline 
$a_2b_2$ & $(-\vt_{b_2}-\a^1\d^1,-\vt_{a_2}-\a^2\d^2,\vt_{b_2}+\vt_{a_2})$ &
$\tilde r_4$ & $\oh(\d^1+\d^2)$\\
$a_2b_2^*$ & $(\vt_{b_2}+\a^1\d^1,-\vt_{a_2}-\a^2\d^2,-\vt_{b_2}+\vt_{a_2})$ &
$\tilde r_4$ & $\oh(-\a^1\d^1+\a^2\d^2)$\\
\hline 
\hline
\end{tabular}
\label{qsusybreaking2}
\caption{\small Small Q-SUSY breaking for a square quiver
model. The specific hierarchy of angles
$\vt_{a_1} < \vt_{a_2} < \vt_{b_1} < \vt_{b_2}$ has been chosen
in order to compute the last column of this table.}}

Is easy to see that if we choose variations $\d^1$, $\d^2$ that satisfy
\beq
\d^1 > \d^2 > \a^1\d^1 > 0 \ {\rm and} \ \a^1 \d^1 > \a^2 \d^2,
\label{deltas}
\eeq
then all scalars appearing at an intersection have positive mass$^2$.
Although we have performed this explicit computation for a specific
hierarchy of angles, we expect this qualitative behaviour to hold in 
general cases. Thus this is a remarkable difference compared to
the $\cn =1$ systems: in Q-SUSY configurations the FI-terms do not
necessarily trigger gauge symmetry breaking
\footnote{Note that SUSY-breaking by FI-terms in
a Q-SUSY model will thus not obey the Ferrara-Girrardello-Palumbo
type of sum-rules \cite{fgp} which precluded the construction
of phenomenological models with tree-level SUSY-breaking
in the early days of SUSY-phenomenology.}. Thus
FI-terms do
actually break $\cn = 1$ SUSY locally (in addition to the overall
SUSY-breaking appearing at the loop level).

\section{Some explicit models}

Starting with the general hexagonal structure discussed in Section 4 one
can construct a variety of interesting Q-SUSY models. In particular, one can
obtain simple models by deleting some of the nodes in the hexagon in
fig.\ref{hexagon}.
We discuss here a couple of examples.

{\it i)  Q-SUSY models from a square quiver}

In specific models not all six types of D6-branes 
$(a_1, b_1, c_1, a_2, b_2, c_2)$
need to be present. For example, one can have a square type of quiver with only
branes of type $a_1, a_2, b_1, b_2$
(see fig.\ref{cuadrado}). The presence of both types of branes $a_1, b_1$ 
and their relatives $a_2, b_2$ which contribute with opposite sign 
to some of the RR tadpoles, allow for the construction
of theses  models  without further addition of other branes 
to cancel tadpoles.

\EPSFIGURE{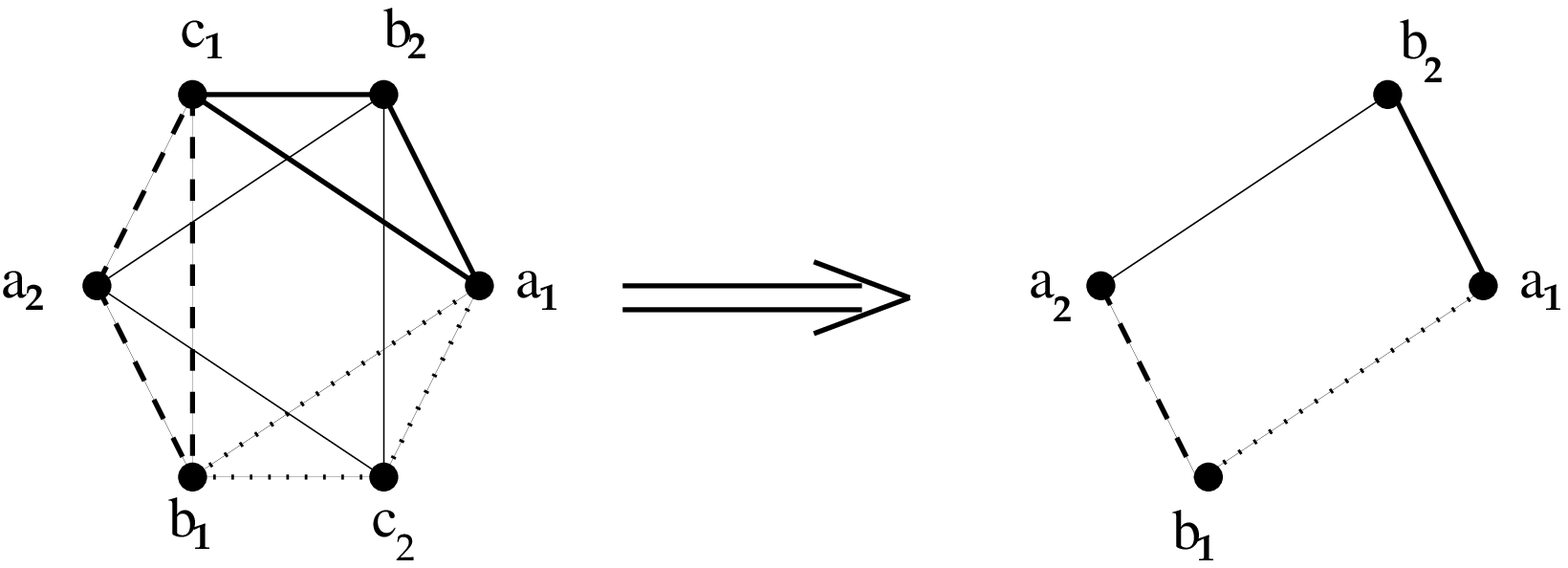, width=5in}
{\label{cuadrado}
Square quiver Q-SUSY model  with four group  
factors. The dots represent the four  different stacks
of branes (plus mirrors)  in the model whereas the links represent the
chiral intersections of those branes. There are four types of links
corresponding to the different SUSY's. }

 In fact such
type of structure is the one appearing in the realistic models  
of \cite{imr} in which the starting gauge group is $U(3)\times U(2)\times
U(1)\times U(1)$. These four group factors correspond to the 
four vertices in the square quiver in fig.\ref{cuadrado}.
 After three  $U(1)$'s get massive through a
generalized Green-Schwarz mechanism, only the SM group survives
and three generations of quarks and leptons
are obtained at the intersections. Although these models
are in general not supersymmetric, it has been recently shown that a
subset of these models has Q-SUSY for appropriate choices
of the complex structure moduli. This class of Q-SUSY models 
yielding the SM is discussed in some detail in a separate paper 
\cite{cim2} and we direct the reader to that reference for more details.

{\it ii) A Q-SUSY Standard Model with $\cn = 1$ 
supersymmetry in the visible sector}

Another interesting class of theories may be obtained starting with three
of the six types of
D6-branes  considered above.
 One can consider theories in which a certain subsector respects a given
$\cn = 1$
supersymmetry. For example, consider three sets of intersecting branes
of type $a_2, b_2, c_2$
(see fig.\ref{triangulo}). The intersections of these three branes respect 
the same SUSY, $\tilde r_4$, thus this subsector of the theory is fully
supersymmetric. In a simple toroidal (non-orbifold) setting as the one
here this cannot be the full story, we already mentioned in Section 3 that
such
a configuration would have RR-tadpoles.
Cancellation of those tadpoles requires extra sources of RR-flux.
In the orbifold models of
\cite{csu} that was achieved by the presence of  three extra
orientifold planes in the system. In our case we will achieve  
that by adding a non-factorizable D6-brane chosen precisely to cancel
the remaining RR-tadpoles
\footnote{Alternatively one can add explicit RR fluxes 
as in e.g. \cite{flujos} in order to cancel tadpoles. 
See in particular \cite{uranga}.}
. Let us call it the H-brane.
It is easy to convince oneself that, in all generality, if the
subsystem of the intersecting branes has an anomaly-free spectrum,
the extra non-factorizable brane H will have no net intersection with
the original (``visible'')
$a_2, b_2, c_2$ brane system. 
Thus, in general fields transforming
both under the ``visible branes'' and the H-brane quantum numbers
will be massive, typically of order the string/compactification scale.
In this way we will have a $\cn = 1$ supersymmetric visible sector formed by
branes of types $a_2, b_2, c_2$ and a sort of ``hidden sector''
 with $\cn = 0$ SUSY in general.

\EPSFIGURE{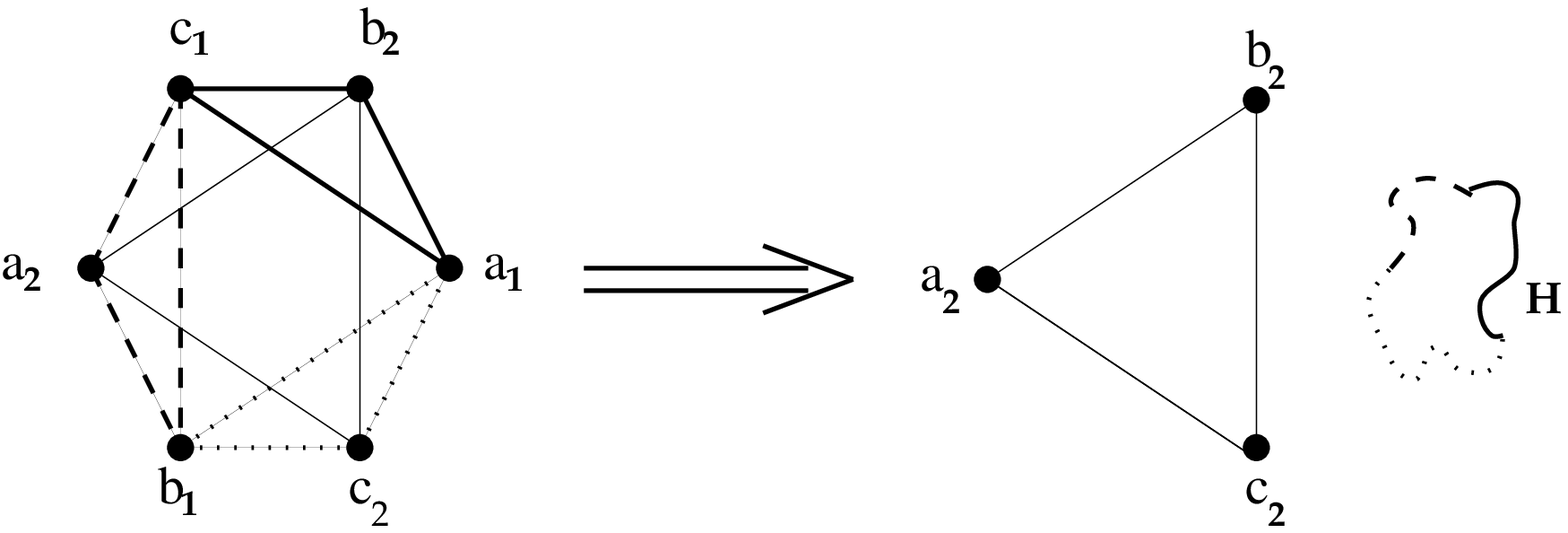, width=5in}
{\label{triangulo}
Model with an anomaly-free  
subsector respecting $\cn = 1$ supersymmetry.
RR tadpole cancellation conditions require in this case the 
presence  of extra sources of RR flux. In particular  
 a non-factorizable brane $H$ can be added which has no intersection
with the ``visible'' $\cn = 1$ SUSY subsector. }

\TABLE{\renewcommand{\arraystretch}{2}
\begin{tabular}{|c|c||c|c|c|}
\hline 
brane\ type  &
 $N_i$    &  $(n_i^1,m_i^1)$  &  $(n_i^2,m_i^2)$   & $(n_i^3,m_i^3)$ \\
\hline\hline   $a_2$
 &  $N_a=3$ & $(1,0)$  &  $(3,1/2)$ &  $(3 ,  -1/2)$  \\
\hline $b_2$ &  $N_b=2$ &   $(1,-1)$    &  $ (2,0)$ & $(1,1/2)$   \\
\hline $c_2$ &  $N_c=1$ & $(0,1)$  & $(0,-1)$  & $(2,0)$  \\
\hline $a_2$' &  $N_d=1$ &   $(1,0)$    &  $(3,1/2 )$ & $(3, -1/2)$   \\
\hline \end{tabular}  
\label{SUSYmodel}
\caption{ Wrapping numbers of a three generation
 SUSY-SM with $\cn = 1$ SUSY locally.}}

\TABLE{\renewcommand{\arraystretch}{1.2}
\begin{tabular}{|c|c|c|c|c|c|c|c|}
\hline Intersection &
 Matter fields  &   &  $Q_a$  & $Q_b $ & $Q_c $ & $Q_d$  & Y \\
\hline\hline $(a_2b_2)$ & $q_L$ &  $2(3, 2)$ & 1  & -1 & 0 & 0 & 1/6 \\
\hline $(a_2b_2^*)$  & $Q_L$   &  $( 3,2)$ &  1  & 1  & 0  & 0  & 1/6 \\
\hline $(a_2c_2)$ & $D_R$   &  $3( {\bar 3},1)$ &  -1  & 0  & 1  & 0 & 1/3 \\
\hline $(a_2c_2^*)$  & $U_R$   &  $3( {\bar 3},1)$ &  -1  & 0  & -1  & 0 & -2/3
\\
\hline ($b_2a_2$') & $ l $    &  $2(1,2)$ &  0   & -1   & 0  & 1 & -1/2  \\   
\hline ($b_2a_2^*$') & $ L $    &  $(1,2)$ &  0   & 1   & 0  & 1 & -1/2  \\
\hline ($c_2a_2$') & $E_R$   &  $3(1,1)$ &  0  & 0  & 1  & -1  & 1   \\
\hline ($c_2a_2^*$') & $N_R$   &  $3(1,1)$ &  0  & 0  & -1  & -1  & 0 \\
\hline ($b_2c_2$) &  $H $    &  $2(1,2)$ &  0   & 1   & -1  &  0 & -1/2  \\
\hline ($b_2c_2^*$) & $ {\bar H} $    &  $2(1,2)$ &  0   & 1   & 1  & 0 & 1/2 \\
\hline 
\end{tabular}
\label{espectrosm}
 \caption{Chiral spectrum of the SUSY SM in the text.}}

\EPSFIGURE{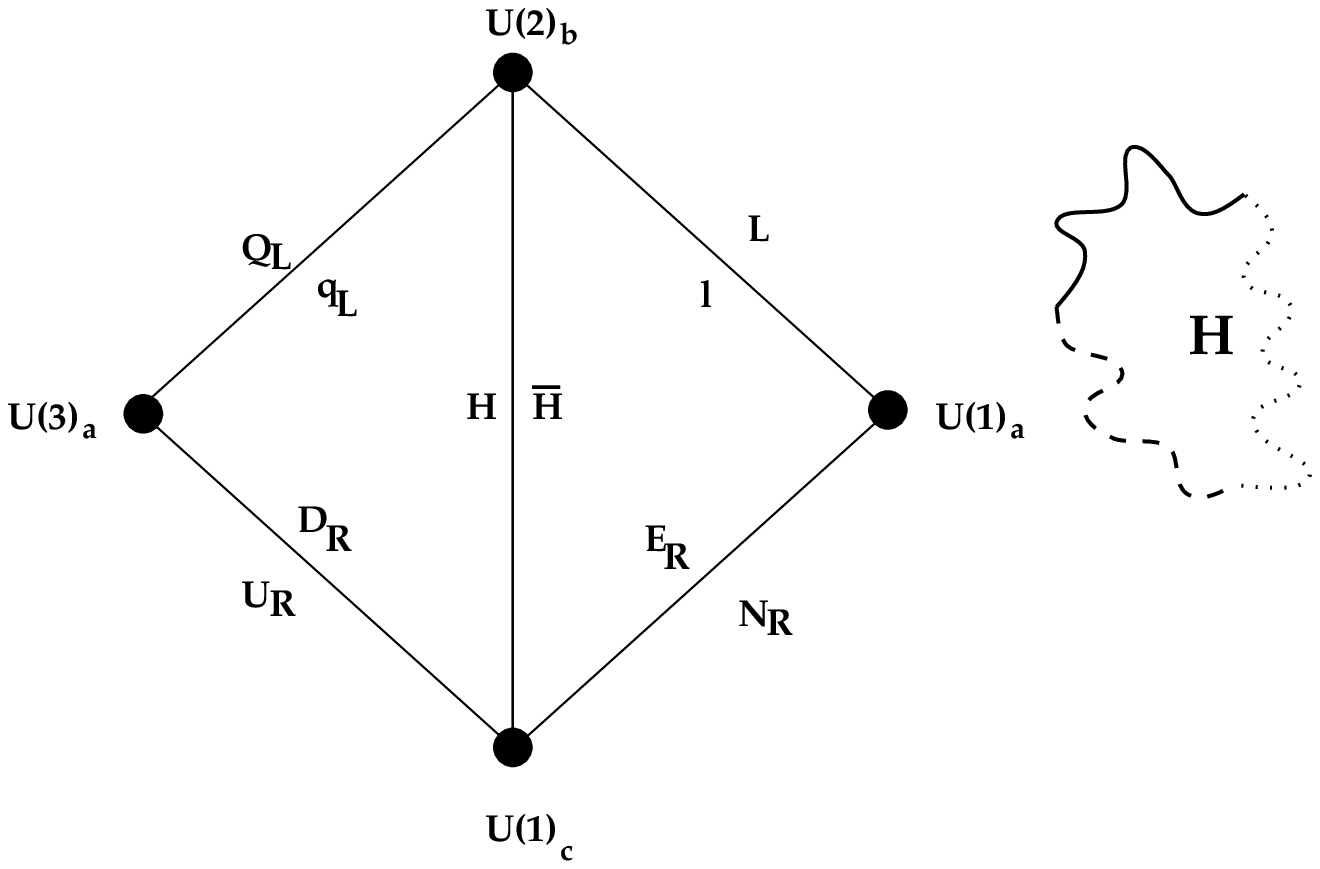, width=5in}
{\label{ssm}
Quiver of the three-generation
SM with $\cn = 1$ SUSY at the local level discussed in the text. }

As an example consider four stacks of branes with wrapping
numbers and multiplicities given in table \ref{SUSYmodel}.
It is easy to check that for appropriate choices of the
tori complex structure there is an unbroken SUSY ($\tilde r_4$)
at all brane intersections,
as we discussed in Section 4. These four sets of branes form
a subsector of the theory with the same  unbroken $\cn = 1$
SUSY. Cancellation of RR-tadpoles requires the presence of
an extra, in general non-factorizable D6-brane $H$ with
vanishing intersection with our
SUSY subsector (see fig.\ref{ssm}). The chiral fields  at each intersection
may be computed and one finds the chiral spectrum in table
\ref{espectrosm}. This spectrum is just the one of the 
Minimal SUSY Standard Model with two Higgs sets.
Although in principle there are four $U(1)$ fields in
the ``visible sector'' two of them are anomalous and become
massive. The two anomaly-free $U(1)$'s are in fact 
$(B-L)$ and the usual hypercharge.

Note that in  these theories the Higgs doublet scalar masses
are protected against quadratic divergences  
up to two loops, as explained in Section 2.
Although the ``visible sector'' of the model has $\cn =1$ SUSY,
there is a non-SUSY massive sector from strings stretching 
between the visible sector and the non-factorizable brane system $H$.
Loops as in fig.\ref{twoloops}-b  will give two-loop contributions to the
Higgs masses  of order $\alpha_2/(4\pi)M_s\sim 3\times 10^{-3}M_s$
\footnote{Interestingly enough,
in case we add explicit RR fluxes to cancel tadpoles
there would be in general no such massive fields 
\cite{uranga} charged 
with respect to the SM interaction. Thus the model would rather look 
like standard $\cn =1$ SUSY models with a SUSY breaking hidden sector.
In this case the $\cn =1$ SUSY of the visible sector would
protect the scalars and we could  then increase
the string scale well above the electroweak scale.}. 
Thus the scale of electroweak scale symmetry breaking 
may be naturally small as long as $M_s\leq 30 TeV$.
On the other hand gaugino masses  will get 
masses at the one-loop level from diagrams as in fig.\ref{gauginos}
\footnote{Note that the structure of the 
low-energy SUSY spectrum would be relatively similar to the
models with gauge mediated SUSY-breaking \cite{gauge}. Thus 
one does not expect important FCNC effects from the
sparticle sector.}.
Electroweak symmetry breaking may proceed in a radiative way
as in the MSSM from the one-loop coupling of Higgs fields to
the heaviest quarks \cite{ir}. 

As will be explained elsewhere  in more detail \cite{cim2} the 
standard model Higgs mechanism 
in intersecting brane models has a nice geometrical interpretation in
terms of recombination of intersecting branes.
Thus e.g. a vev for the Higgs fields $H$ and ${\bar H}$ would
correspond to the recombination of three branes:
\beq
b_2 \ +\ {b_2^*}^\prime \ +\ c_2 \ \ \rightarrow \ \  f
\label{recomb}
\eeq
into a single one $f$. Here $b_2$ and $b_2'$ denote the two
branes giving rise to the $U(2)_L$ gauge interactions.
This recombination proceeds by a smoothing out of the intersections,
which is controlled by the vevs of the Higgs fields.
In the final configuration there are only three stacks of branes
$a_2$,$a_2'$ and $f$ and the gauge group is $SU(3)_c\times U(1)_{em}$
\footnote{Actually in this particular example there is an extra 
unbroken $U(1)$ related to $B-L$ which was already present in the
initial configuration. The Higgs fields are neutral under it and do not
give it a mass. It may become massive if a right-handed  sneutrino mass 
gets a vev \cite{cim2}. A third $U(1)$ gets massive from the presence of
$B\wedge F$ couplings.}.
Once the Higgs fields get a vev, the quarks and leptons
in the standard model get masses. In the language of
brane recombination this can be verified by noting that 
the final recombined brane $f$ has no intersection with the 
other two, no chiral fermions are left. Thus for example,
$I_{a_2f}=I_{a_2b_2}+I_{a_2b_2^*}'+I_{a_2c}=2+1-3=0$, there are
no chiral quarks left.

\section{Final comments and conclusions}

In this paper we have presented a  class of $D=4$ chiral
field  theories with interesting loop stability properties.
They can be depicted as quivers in which we have extended   
SUSY gauge theories at the nodes and  bifundamental chiral  
multiplets with respect to different $\cn =1$ SUSY's at the 
links.   Altogether the theories are not supersymmetric but the
scalars only feel that SUSY has been broken at two loops.
This property may be interesting in order to understand the
stability of a hierarchy between a fundamental scale $\sim $
10-100 TeV and the weak scale of order 0.1 TeV.

It is important to realize that  some level of 
low-energy supersymetry 
may be already needed to understand the precision
LEP data \cite{barbieri}. This need is independent  
of the solution proposed for the classical gauge hierarchy 
problem between the weak scale and the Planck scale.
Even alternative solutions like a low string scale 
slightly above the weak scale have eventually 
to face this modest fine-tuning problem. 
On the other hand full $\cn = 1$ supersymmetry is 
more than what we actually need in models with
low string scale. In this context  
Q-SUSY models may be of phenomenological relevance.

Field theories with these properties naturally appear in
toroidal compactifications of Type II string theory
with intersecting D6-branes wrapping 3-cycles in the  
torus. For particular choices of the tori radii
one gets chiral SUSY multiplets at the different intersections,
but generically preserving different $\cn =1$ subgroups,
yielding the searched  Q-SUSY structure.
We have studied certain aspects of the effective Lagrangian
at those intersections, which apply both to
complete $\cn = 1$ SUSY configurations and the Q-SUSY configurations
here introduced. In particular we compute the gauge coupling constants 
from the Born-Infeld action and from the effective Lagrangian point
of view (using holomorphicity and anomalous RR-couplings)
and show their agreement. We also study small perturbations
of the complex structure moduli away  from the SUSY point.
We observe  that, as expected, this produces SUSY-breaking by FI-terms. We
compute those FI-terms
and show how the masses given to scalars coincides with the computation
from string mass formulae.

In Section 6
we construct several explicit D-brane models.
In particular one can construct theories
which subsectors respecting  $\cn = 1$ locally but in
which massive $\cn = 0$ sectors induce
SUSY-breaking in loops. A particular interesting model is constructed
with a massless spectrum strikingly similar to the MSSM.
This gives us an explicit  example of a D-brane theory in which
the Higgs mass is stable up to two loops, providing
a stable hierarchy between a string scale of order 30 TeV
and the weak scale. The usual Higgs mechanism has a nice
geometrical interpretation in terms of brane recombination: the
branes of the electroweak group recombine into a single 
brane which is related to electromagnetism. At the
same time the quarks and leptons become massive in the process.
We leave more phenomenological aspects of this and other    
explicit models for a separate paper \cite{cim2}.

One point we have not addressed explicitly in this paper
is the stability of the considered
specific D6-brane configurations.
The models are non-supersymmetric and hence NS-NS tadpoles
are expected, which show an instability of the theory
{\it in a flat background}.
 These may be understood as coming from
the existence of a non-vanishing tree-level scalar
potential for the real part of the
 moduli $S, U^I$. The Q-SUSY models turn out
to have a particularly simple structure for this potential  
compared to generic toroidal constructions. The potential is
linear in those NS fields and in particular cases some of the NS tadpoles
(but not all) can be shown to cancel if the corresponding RR
tadpole does. On the other hand one may perhaps get rid of the
instability a la Fischler-Susskind \cite{fs}, by redefining the closed
string background. This redefinition has been shown to lead to
warped metrics in some particular simple examples \cite{Dudas}.
In realistic models to study the viability of such a
procedure may be quite complicated.

Other related issue is that
of obtaining adequate $D=4$ gravity, with a large Planck scale
compared to the string scale in the toroidal D-brane constructions.
One cannot get a large Planck scale by
taking a large transverse volume, because in the
models considered there is not a volume which is transverse
to all the branes simultaneously. In this context perhaps
a gravity localization a la Randall-Sundrum could be
at work. The presence of warping factors from a non-vanishing
NS scalar potential could be relevant if that possibility is present.

Notice finally that
the class of Type II orientifold configurations here studied have 
$\cn = 4$ supergravity in the bulk ($\cn = 8$ in the case of toroidal
compactifications). The SUSY partners of the graviton will only
get their mass from the SUSY-breaking effects on the branes.   
It would be interesting to study possible consequences of such
approximate extended supergravity in the bulk
(see e.g. \cite{schmidhuber} and references therein).

\bigskip

\bigskip

\centerline{\bf Acknowledgements}

\bigskip

We are grateful to G. Aldaz\'abal, F. Quevedo,
R. Rabad\'an and  A. Uranga  for very
useful  comments and discussions.
The research of D.C. and F.M. was supported by
 the Ministerio de Educacion, Cultura y Deporte (Spain) through FPU grants.
This work is partially supported by CICYT (Spain) and the
European Commission (RTN contract HPRN-CT-2000-00148).


\end{document}